\RequirePackage{fix-cm}
\documentclass[smallextended,natbib]{svjour3}       % onecolumn (second format)
\smartqed  % flush right qed marks, e.g. at end of proof
\usepackage[utf8]{inputenc}
\usepackage[T1]{fontenc}
\usepackage{microtype}

\usepackage{graphicx}

\usepackage{newtxtext,newtxmath} % replacement of mathptmx, use Times fonts if available on your TeX system

\usepackage{color, colortbl}
\usepackage{tcolorbox}
\usepackage{xcolor}
\usepackage{tabularx}
\usepackage{amsmath,amsfonts}
\usepackage[noend]{algpseudocode}

\usepackage{booktabs}
\usepackage{multirow}
\usepackage{epstopdf}
\usepackage{tikz}
\usepackage{enumitem}
\usepackage{listings}
\usepackage[linesnumbered,ruled,lined]{algorithm2e}
\usepackage{alltt}
\usepackage{multirow}
\usepackage{textcomp}
\usepackage{framed}
\usepackage{hhline}

\usepackage{caption} % For customizing captions
\usepackage[skip=4pt]{subcaption}

\usepackage{array}

\usepackage{flushend}
\usepackage{makecell}
\usepackage{balance}
\usepackage{pifont}
\usepackage{centernot}
\usepackage{mathtools}
\usepackage{stmaryrd}

\usepackage[colorinlistoftodos,textwidth=15mm,textsize=tiny,obeyFinal]{todonotes}
\usepackage{lstautogobble}
\usepackage{threeparttable}
\usepackage{stmaryrd}
\SetSymbolFont{stmry}{bold}{U}{stmry}{m}{n}

\usepackage{fancyvrb}
\usepackage{adjustbox}

\usepackage{svg}
\svgsetup{inkscapepath=svgsubdir}

\usepackage{cuted}

\usepackage{tabularray}
\NewTblrTheme{mockcaption}{
\SetTblrStyle{caption-tag}{font=\normalsize}
\SetTblrStyle{caption-sep}{font=\normalsize}
\SetTblrStyle{caption-text}{font=\normalsize}
\SetTblrOuter{headsep=4pt}
\SetTblrTemplate{caption}{simple}
\SetTblrTemplate{capcont}{simple} % 若为跨页表格的延续标题也一并设置
\SetTblrStyle{note}{font=\footnotesize}
\SetTblrStyle{remark}{font=\footnotesize}
}

\usepackage[
    colorlinks, % 关键：开启这个，方框就会消失
    breaklinks, % 允许链接在行末折断，且保持跳转完整性
    allcolors=black, % 先把所有链接设为黑色（包括图表引用）
    citecolor=blue,
    urlcolor=blue
]{hyperref}

\usepackage{fontawesome5} % fontawesome

\newcommand{\sootup}{\textsc{SootUp}\xspace}
\newcommand{\soot}{\textsc{Soot}\xspace}
\newcommand{\sbugs}{\textsc{SpotBugs}\xspace}
\newcommand{\fbugs}{\textsc{FindBugs}\xspace}
\newcommand{\alrf}{\textsc{ANTLR4}\xspace}
\newcommand{\alrt}{\textsc{ANTLR3}\xspace}
\newcommand{\sversion}{4.5.0}
\newcommand{\stversion}{1.3.0}
\newcommand{\sbversion}{4.5.0}
\newcommand{\fbversion}{3.0.1}
\newcommand{\alrtversion}{3.5.3}
\newcommand{\alrfversion}{4.13.2}

\newcommand{\ie}{{i.e.},\xspace}
\newcommand{\eg}{{e.g.},\xspace}

\newcommand{\prsubmit}{10\xspace}
\newcommand{\issuesubmit}{five\xspace}
\newcommand{\prmerged}{eight\xspace}

\newcommand{\numprks}{27\xspace} % number of PRs by keyword searching
\newcommand{\numcodeks}{22\xspace} % number of code instances by keyword searching
\newcommand{\numdocks}{seven\xspace} % number of documents by keyword searching
\newcommand{\numcmtks}{13\xspace} % number of commits by keyword searching
\newcommand{\numisks}{30\xspace} % number of issues by keyword searching
\newcommand{\numcpcode}{17\xspace} % number of code instances by keyword searching: Copy old code into comments

\newcommand{\stlabeltotal}{37\xspace}  % sootup label search: total issues
\newcommand{\stlabelbug}{16\xspace}  % sootup label search: issues labeled as bug
\newcommand{\stnolabel}{21\xspace}   % sootup label search: issues without label

\newcommand{\numdataset}{1,805\xspace}

\definecolor{orcidlogocol}{HTML}{A6CE39}
\newcommand{\orcid}[1]{\href{https://orcid.org/#1}{\textcolor{orcidlogocol}{\faOrcid}}}

\usepackage[misc]{ifsym}

\newtcbox{\mybox}[1][breakable]{on line, enlarge top by=10pt, enlarge bottom by=10pt,
     boxsep=8pt, boxrule=2pt, size=small, arc=1mm}

\newcommand{\nbc}[3]{
	{\colorbox{#3}{\bfseries\sffamily\scriptsize\textcolor{white}{#1}}}
	{\textcolor{#3}{\sf\small$\blacktriangleright$\textit{#2}$\blacktriangleleft$}}
}

\definecolor{grey}{rgb}{0.7,0.7,0.7}

\newcommand{\lstbg}[3][0pt]{{\fboxsep#1\colorbox{#2}{\strut #3}}}
\lstdefinelanguage{diff}{
  basicstyle=\ttfamily\scriptsize,,
  morecomment=[f][\lstbg{red!20}]-,
  morecomment=[f][\lstbg{green!20}]+,
  morecomment=[f][\lstbg{yellow!20}]++,
  morecomment=[f][\textit]{@@},
  texcl=false
}

\definecolor{todocolor}{rgb}{0.9,0.1,0.1}
\definecolor{indiagreen}{rgb}{0.07, 0.53, 0.03}
\definecolor{hycolor}{rgb}{0.7,0.7,0.3}
\definecolor{darkbrown}{rgb}{0.4, 0.26, 0.13}

\tikzstyle{highlighter} = [
  yellow,
  line width = \baselineskip,
]

\newcounter{highlight}[page]
\newcommand{\tikzhighlightanchor}[1]{\ensuremath{\vcenter{\hbox{\tikz[remember picture, overlay]{\coordinate (#1 highlight \arabic{highlight});}}}}}

\newcommand{\ignore}[1]{}

\newcommand{\findingcnt}{eight}

\definecolor{main-color}{rgb}{0.6627, 0.7176, 0.7764}
\definecolor{string-color}{rgb}{0.3333, 0.5254, 0.345}
\definecolor{key-color}{rgb}{0.8, 0.47, 0.196}
\lstdefinestyle{mystyle} {
    language = Java,
    basicstyle = {\ttfamily \color{main-color}},
    stringstyle = {\color{string-color}},
    keywordstyle = {\color{key-color}},
    keywordstyle = [2]{\color{lime}},
    keywordstyle = [3]{\color{yellow}},
    keywordstyle = [4]{\color{teal}},
    morekeywords = [3]{<<, >>},
    morekeywords = [4]{++},
    basicstyle=\ttfamily\scriptsize,
    commentstyle=\color{blue}\ttfamily,
    morecomment=[f][\lstbg{red!20}]-,
    morecomment=[f][\lstbg{green!20}]+,
    morecomment=[f][\lstbg{yellow!20}]++,
    morecomment=[f][\lstbg{yellow!20}]--,
    morecomment=[f][\textit]{@@},
    breaklines=true,
    texcl=false
}
\journalname{Empirical Software Engineering}
\begin{document}

% \makeatletter
% \typeout{NORMALSIZE = \f@size pt}
% \small
% \typeout{SMALL = \f@size pt}
% \footnotesize
% \typeout{FOOTNOTESIZE = \f@size pt}
% \scriptsize
% \typeout{SCRIPTSIZE = \f@size pt}
% \makeatother

\title{Bidirectional Code Reuse in Software Redesign: An Action Research Study of Static Analyzers}
% \subtitle{Do you have a subtitle?\\ If so, write it here}


\author{Xiaowen Zhang \orcid{0009-0009-4475-9442} \and 
        Huaien Zhang \orcid{0000-0001-6498-5062} \Letter \and 
        Shin~Hwei~Tan \orcid{0000-0001-8633-3372}}

\authorrunning{X. Zhang et al.} % Short form of author list, if too long for running head

\institute{Xiaowen Zhang \and Shin Hwei Tan \at
              Concordia University, Montreal, Quebec, Canada \\
              \email{xiaowen.zhang@mail.concordia.ca, shinhwei.tan@concordia.ca} 
           \and
           Huaien Zhang \at
              The University of Hong Kong, Hong Kong, China \\
              \email{huaien@hku.hk}
}

\date{Received: date / Accepted: date}

\maketitle

\begin{abstract}
Software redesign preserves functionality while improving quality attributes, but manual reuse of code and tests is costly and error-prone, especially in cross-repository redesigns. Focusing on static analyzers where cross-repository redesign needs often arise, we conduct a bidirectional study of the ongoing \soot/\sootup{} redesign using an action research methodology that combines empirical investigation with validated open-source contributions. Our study reveals: (1) non-linear migration that necessitates bidirectional reuse, (2) deferred reuse via TODOs, (3) neglected test porting, and (4) residual bug propagation during migrations. We identify tracking corresponding code and tests as the key challenge and address it by retrofitting clone detection to derive code mappings between original and redesigned projects. Guided by semantic reuse patterns derived from our study, we propose the Semantic Alignment Score (SAS), which incorporates semantic cues from preserved identifiers, API documentation, and comments. Evaluations on three redesigned project pairs (\soot/\sootup, \fbugs/\sbugs, and \alrt/\alrf) show that SAS improves average F1 for code mapping detection by up to 0.34 on our manually labeled benchmark of \numdataset{} method pairs, while strong traditional detectors approach or surpass LLM detectors. The code mapping analysis also uncovered ongoing maintenance needs, leading to \issuesubmit{} issues and \prsubmit{} pull requests, of which \prmerged{} have been merged.

% Software redesign preserves functionality while improving quality attributes, but manual reuse of code and tests is costly and error-prone, especially in cross-repository redesigns. We study the ongoing \soot/\sootup{} redesign and find that code reuse is bidirectional and adapts to the new context, introducing divergence while preserving substantial textual similarity to the original code. Existing clone detectors, designed primarily around structural and syntactic similarity, struggle to distinguish code mappings between original and redesigned projects from irrelevant clones, particularly in test code where shared test-harness APIs make unrelated test cases appear similar. Guided by these findings, we propose the Semantic Alignment Score (SAS), which retrofits existing clone detectors to support code mapping detection using semantic cues from preserved identifiers, API documentation, and comments. We evaluate SAS on a manually labeled benchmark of \numdataset{} method pairs across three redesigned project pairs (\soot/\sootup, \fbugs/\sbugs, and \alrt/\alrf). SAS improves average F1 for code mapping detection by up to 0.34 on this benchmark, enabling strong traditional detectors to approach or surpass the effectiveness of LLM detectors without LLM inference costs. The code mapping analysis also uncovered ongoing maintenance needs, leading to \issuesubmit{} issues and \prsubmit{} pull requests, of which \prmerged{} have been merged.


\end{abstract}

\section{Introduction}
\label{sec:intro}

Software redesign (or rewrite) is a challenging maintenance task that improves non-functional properties (e.g., extensibility) while preserving behavior~\citep{stuurman2016analyzing}.
During software redesign, developers must balance maintaining core functionality with major structural improvements. 
Ensuring functional equivalence often requires manually reusing code and tests, which is costly and error-prone. 
% Maintaining functional continuity often involves manually reusing relevant code and tests from the original project in the redesigned project, which can be costly and error-prone. 
To make such cross-project reuse traceable, we formalize the relationships between corresponding code as \emph{code mappings}.
% Ensuring functional equivalence often requires manually reusing code and tests, which is costly and error-prone. 

\noindent\textbf{Why Static Analyzers?}
Designing static analyzers that are both precise and efficient is inherently challenging, as architectural changes often ripple through program representations, APIs, and performance-critical components~\citep{10.1007/978-3-319-89884-1_23}, making redesign unavoidable. Consequently, the need for software redesign often arises in  static analyzers.
For example, \soot~\citep{lam2011soot,vallee2000optimizing,park2022android,arzt2017soot,arzt2013instrumenting,arzt2014flowdroid,zhang2024sascope}, a widely used static analysis framework was redesigned as \sootup in 2022 to address long-standing architectural limitations~\citep{karakayasootup}, yet the migration remains incomplete after more than three years, requiring concurrent maintenance of both projects. A similar redesign occurred when  \fbugs~\citep{findbugslink} was succeeded by \sbugs~\citep{findbugsdead}, and when \alrt~\citep{antlr3github,antlr3paper2011} was rewritten as \alrf~\citep{antlr4github,antlr4paper2014}, replacing its parsing algorithm entirely.
The ongoing \soot/\sootup redesign provides us with a unique opportunity to conduct an action research study of bidirectional code reuse, whereas the completed \fbugs/\sbugs and \alrt/\alrf migration allows us to evaluate the generalizability of our proposed technique.

These redesigns differ fundamentally from commonly studied redesign scenarios---such as mobile application UI migrations that typically occur within a single repository and are traceable via commit histories (e.g., AntennaPod~\citep{antennapod} migrated from Material 2 to Material 3) ~\citep{androidui2022,user_centered_2023,fuada2024ui,androidreview2025}.  In contrast, static analyzer redesigns often span multiple repositories, involve substantial architectural changes, and evolve as independent projects. As a result, the evolution history that would normally connect corresponding code through version control is fragmented across repositories, making functional correspondence difficult to trace.
%These characteristics complicate reasoning about functional correspondence across versions.

To understand how such redesigns unfold in practice, we conduct a bidirectional study of the ongoing \soot/\sootup redesign, mining redesign-related changes from software artifacts (e.g., issues and pull requests (PRs)) via cross-references between the two projects. This bidirectional perspective is necessary because developers often modify both projects concurrently.

Our study identifies a recurring challenge: \emph{developers struggle to track corresponding production and test code as the original and redesigned projects evolve}. While reused fragments across the two projects often resemble code clones, only a subset of these clones reflect meaningful functional migration relationships. Distinguishing these relationships proved difficult in practice and hindered reasoning about redesign progress and correctness.

\begin{figure}[htbp]
    \centering
    \includegraphics[width=0.4\textwidth]{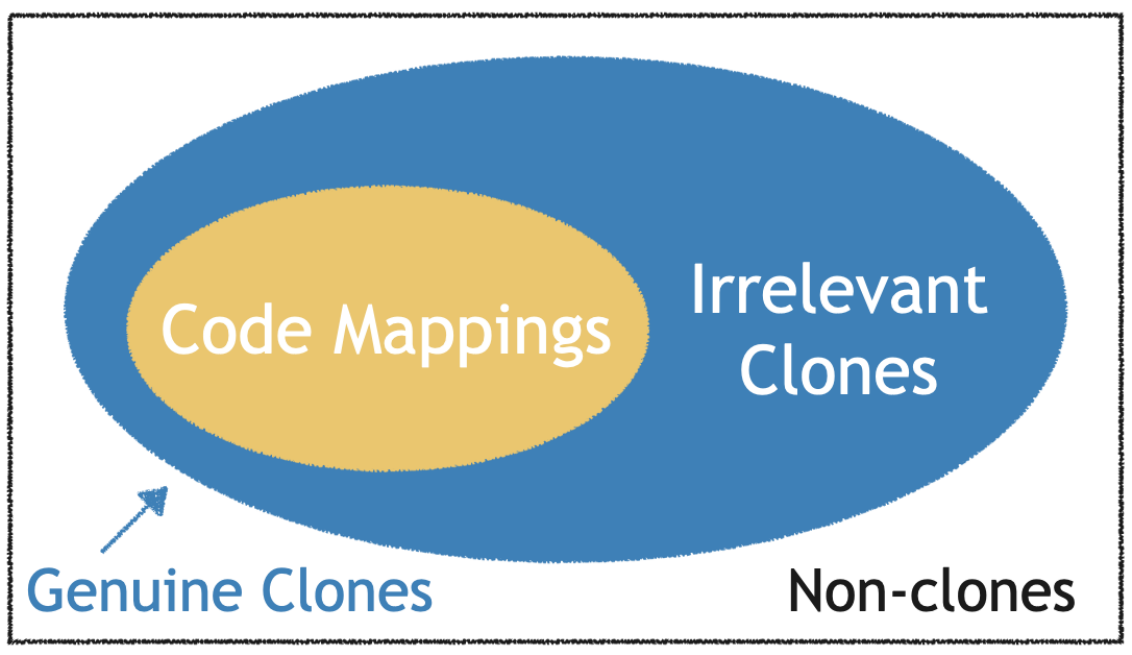} 
    \caption{Relationship between non-clones, genuine clones, code mappings and irrelevant clones in $D$.}
    \label{fig:relationship}
\end{figure}

Motivated by this challenge, we explored whether clone detection techniques could be adapted to 
identify \emph{code mappings} between the original and redesigned projects, expressed as method pairs. 
We represent potential mappings as method pairs (Figure~\ref{fig:relationship}), and categorize them into genuine clones, code mappings, irrelevant clones, and non-clones, as described below: 

\noindent\textbf{Genuine Clone (GC)} denotes a method pair ($m_1$, $m_2$) satisfying the traditional clone definitions (Type-1 to Type-4, see Section~\ref{sec:background})~\citep{svajlenko2015evaluating,wang2023ccstokener}; 

\noindent\textbf{Code Mapping (CM)} refers to a method pair ($m_1 \in proj_{orig}$, $m_2 \in proj_{redesign}$) representing a functional migration link where $m_2$ is either the direct one-to-one functional counterpart of $m_1$ after migration, or transitively mapped to $m_1$ via a Type-1 within-project clone. 

\noindent\textbf{Irrelevant Clone} denotes any genuine clone that is not a code mapping, representing the noise that hinders precise code mapping identification.

\noindent\textbf{Non-clone} denotes all method pairs that fail to satisfy the GC criteria.

We evaluated four representative method-level clone detection techniques: three traditional techniques (NiCad, DeepSim, and CCSTokener) and one LLM-based approach (instantiated with GPT-OSS-120B and Qwen2.5-Coder-32B-Instruct)~\citep{roy2008nicad,zhao2018deepsim,wang2023ccstokener,llm4ccd}. We evaluated all techniques on three redesigned project pairs: the ongoing \soot/\sootup redesign and two completed cases, \fbugs/\sbugs and \alrt/\alrf. 
Our evaluation revealed three key challenges: 
\textbf{(CH1)} traditional techniques often produce many \emph{false positives (FPs)}, where the identified code pairs share low semantic similarity but are incorrectly  marked as clones due to their structural or syntactic similarities (e.g., tokens, AST paths); 
\textbf{(CH2)} traditional techniques lack support for distinguishing meaningful code mappings from irrelevant clones. 
\textbf{(CH3)}  LLM-based approaches~\citep{llm4ccd} have limited scalability due to costly pairwise comparisons~\citep{unseenFuncClones}. 
These challenges motivate our primary goal: improving the precision of traditional detectors to approach the effectiveness of LLM detectors.

To address the aforementioned challenges, we enhance prior clone detection techniques with a set of rules that prioritize preserved natural language information (e.g., identifiers, documentation, and comments) across redesigned projects.
We call these rules 
\textit{Semantic Alignment Heuristics (SAH)} that capture preserved semantic cues (e.g., identifiers, comments) across original and redesigned projects. 
To address \textit{CH1} and \textit{CH2},
we compute the \textit{Semantic Alignment Score (SAS)} for each method pair to filter out unlikely or irrelevant pairs, improving precision while maintaining coverage. To support repository-scale analysis for LLM-based approaches (address \textit{CH3}), 
we further designed a hierarchical detection pipeline that performs class-level pre-filtering prior to method-level analysis, substantially reducing computational and monetary costs.

\begin{table}[!t]
\caption{Empirical findings overview}
\label{tab:findings}
\small
\begin{tblr}{
  colspec = {X[1.1,l,p]X[2.2,l,p]Q[l,p]},
  hlines,
  vlines,
  row{1} = {bg=gray!30, font=\bfseries},
  cells = {valign=t},
}
Empirical theme & Key finding & Details \\
Reuse characteristics in software redesign &
Related changes during redesign are fragmented across multiple software artifacts, while code reuse is bidirectional (involving both forward porting and backporting) and adapts to the new context, often reducing textual similarity. Together, these characteristics make cross-project reuse relationships difficult to track. &
Findings 1--3 \\
Limitations of existing clone detectors &
Traditional detectors produce substantial false positives in both production and test code, with test harnesses introducing additional clone noise in test code. They also tend to be less effective than LLM detectors at identifying code mappings. &
Findings 4, 6 \\
Effectiveness of redesign-aware filtering &
Filtering candidate pairs using preserved identifiers and natural-language information substantially improves traditional clone detectors, with gains of up to 0.34 in average F1 and LLM-comparable CM detection effectiveness for strong traditional detectors. &
Findings 7, 9  \\
\end{tblr}
\end{table}

Overall, this paper makes the following contributions:
\begin{itemize}
    \item \textbf{Action Research Study:} We conduct the first bidirectional study on software reuse in redesigned projects from the perspective of static analyzers using an action research methodology~\citep{runeson2009guidelines,avison1999action} where we identified issues in the first \emph{exploratory stage}, followed by the \emph{action stage} where we brought about changes by \emph{making carefully reviewed and validated open-source contributions} to address the issues; in the \emph{final stage}, we reflected upon the code reuse practice. 
    Our study derives \findingcnt key findings that characterize production and test code reuse in cross-repository redesign. Table~\ref{tab:findings} highlights the key takeaways, with complete findings presented in Sections~\ref{sec:study} and~\ref{sec:eval}. Our contributions to the ongoing \soot/\sootup redesign include:
    reporting four bugs (three fixed), backporting three fixes with tests, and forward porting three validators to the redesigned project.
   
    \item \textbf{Redesign code-mapping dataset:} We contribute a \textit{code mapping dataset for redesigned projects}, comprising \numdataset manually labeled method pairs from three redesigned project pairs, covering production and test code. Pairs are categorized as genuine clones (Type-1--4), code mappings, or non-clones. The dataset provides a foundation for studying code mappings in real-world redesigns, where such labeled data are scarce.

    \item \textbf{Redesign-aware clone filtering:} 
    Guided by empirical reuse patterns, we propose redesign-aware heuristics that retrofit existing traditional clone detectors to establish code mappings between original and redesigned projects. Our approach leverages redesign-specific renaming, preserved identifiers, and natural-language cues to remove the majority of irrelevant clone candidates. This improves average F1 by up to 0.34, enabling traditional detectors to approach or surpass LLM detectors.

\end{itemize}

\section{Background and Tool Selection}
\label{sec:background}

\subsection{Redesigned Projects}
We introduce three redesigned project pairs: \soot/\sootup, \fbugs/\sbugs and \alrt/\alrf, and justify their selection as research targets.

\noindent\textbf{\soot.} \soot is a popular static analysis framework widely used to analyze, instrument, and optimize Java and Android applications~\citep{lam2011soot,vallee2000optimizing,park2022android,arzt2017soot,arzt2013instrumenting,arzt2014flowdroid}. Sponsored by Amazon Web Services, \soot is an open-source project with over 2,000 stars on GitHub.
 
\noindent\textbf{\sootup.}
In December 2022, \sootup was released as a new version of \soot with a completely redesigned architecture.
The project aims to restructure \soot away from heavy use of singletons, providing a lighter and more easily embeddable library.
It also emphasizes parallelization to improve execution speed~\citep{karakayasootup}.
Compared to its predecessor \soot, \sootup includes~\citep{sootuprepo}: (1) improved APIs without globals/singletons, (2) parallelizable architecture, (3) lazy class loading,
(4) fail-early strategy (input validation during object construction), and (5) new source code frontend. Similar to \soot, \sootup is open-source with over 700 stars on GitHub.

\noindent\textbf{\fbugs.}
\fbugs is an open-source static analyzer for detecting bugs in Java programs~\citep{findbugslink} that analyzes Java bytecode.

\noindent\textbf{\sbugs.} \sbugs is the spiritual successor of \fbugs, carrying on from the point where it left off with the support of its community. In 2016, the lead of \fbugs became inactive, leaving numerous community issues unresolved. This prompted volunteers to create a modernized, more maintainable project for Java. On September 21, 2017, the first official version of \sbugs v3.1.0 was released. Unlike \fbugs, it supports language features of newer versions of Java. 

\noindent\textbf{\alrt.} \alrt is a widely used parser generator for constructing lexers, parsers, and tree walkers from grammar specifications, and has been used to build compilers, translators, and program analysis tools~\citep{antlr3github,antlr3paper2011}.

\noindent\textbf{\alrf.} \alrf is a major redesign of \alrt maintained in a separate repository, with substantial changes to the parser generator's implementation and APIs~\citep{antlr4paper2014}. \alrf is open-source with over 19K stars on GitHub~\citep{antlr4github}. We focus on the implementation of the ANTLR tool and its tests, where the primary redesign occurs, excluding language-specific runtime libraries.

\noindent\textbf{Selection of redesigned projects.}
We select \soot/\sootup, \fbugs/\sbugs, and \alrt/\alrf as the target project pairs based on their popularity,and open-source availability, using versions \sversion/\stversion, \fbversion/\sbversion, and \alrtversion/\alrfversion, respectively.
First, they have been extensively studied in domains such as automated testing, software evolution, and bug detection~\citep{zhang2023efficient,tomassi2018bugs,wang2022find,lavazza2020empirical,zhang2023statfier,zhang2024annatester}.
Second, these projects are important and popular because \emph{they are developed and widely used by software engineering and programming language researchers to build automated tools} (i.e., many projects~\citep{lam2011soot,vallee2000optimizing,park2022android,arzt2017soot,arzt2013instrumenting,arzt2014flowdroid} rely on them and need updates to adopt the redesign versions).
Third, their open-source status enables a comprehensive analysis of source code, issues, and PRs. 

\subsection{Code Clone and Detection Tool}
\label{introclone}
Code clones are typically classified into four types~\citep{svajlenko2015evaluating,wang2023ccstokener}: \textbf{Type-1} (exact clones) are syntactically identical, ignoring whitespace and comments. \textbf{Type-2} extend this to include variations in identifier names and literal values. \textbf{Type-3} (near-miss clones) involve structural differences like added, modified, or removed statements. Though functional similarity is not required, their syntactic resemblance often yields similar semantics. In contrast, \textbf{Type-4} are syntactically dissimilar but semantically equivalent (\eg the example in Figure~\ref{fig:problem-clones}). 

\noindent\textbf{Selection of Clone Detection Tools.}
Analyzing code clones is a common approach to studying code reuse.
Several code clone detection techniques~\citep{ccdreview2019} have been proposed. 
Our study focuses on five techniques representing diverse algorithms: NiCad~\citep{roy2008nicad} (text-based), DeepSim~\citep{zhao2018deepsim} (graph- and learning-based), CCStokener~\citep{wang2023ccstokener} (token-, tree-, and graph-based), SSCD~\citep{sscd2022} (embedding-based retrieval) and an LLM-based technique~\citep{llm4ccd}. We selected these tools for their strong performance compared to other clone detectors~\citep{wang2023ccstokener,wang2018ccaligner,amain2022,saini2018oreo} and their support for method-level granularity--a commonly used unit for API documentation and unit testing. 

%(see the supplementary material for our selection criteria and reasons for excluding certain tools). 
%% Our study focuses on four method-level tools representing diverse algorithms: NiCad~\citep{roy2008nicad} (text-based), DeepSim~\citep{zhao2018deepsim} (learning-based), CCStokener~\citep{wang2023ccstokener} (token-based) and an LLM-based technique~\citep{llm4ccd}. 
% Additionally, we use Deckard~\citep{Deckard} to find exact clones in Section~\ref{sec:clone-prevalence}. Deckard is a block-level code clone detection tool based on a tree similarity algorithm. 
% % NiCad is a text-based tool designed to detect clones of comparable size, supporting both function and block granularities. DeepSim employs a graph-based and learning-based approach. We selected NiCad and DeepSim due to their favorable efficacy compared to other clone detection tools~\citep{wang2023ccstokener,wang2018ccaligner,amain2022,saini2018oreo} and their support for function granularity. We select CCStokener as it utilizes a hybrid methodology that integrates token-, tree-, and graph-based techniques.
% Additionally, we use Deckard~\citep{Deckard}, a block-level clone detection tool based on a tree similarity algorithm, to identify exact clones in Section~\ref{sec:clone-prevalence}.

% % \input{table-st-candidate-tools}

% \section{Reuse Characteristics and Patterns: A bidirectional study in \soot and \sootup}
\section{Reuse Characteristics and Patterns}
\label{sec:study}

% To understand how developers perform code and test reuse, we conducted a bidirectional study to answer the question below:
% \noindent\textbf{RQ0:} 
% \textit{What unique characteristics emerge in production and test code reuse between the original and redesigned projects (\soot vs. \sootup)?}

\begin{figure}[!hbt]
    \centering
    \includegraphics[width=\textwidth]{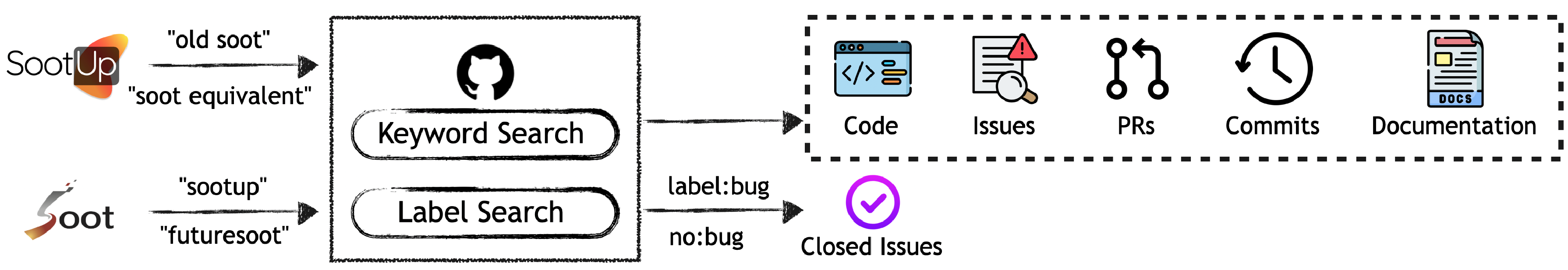}
    \caption{Overview of Our Bidirectional Study of an Ongoing Redesigned Project}
\label{fig:bi-overview}
\end{figure}

To understand how developers perform code and test reuse, we conducted a bidirectional study of the unique characteristics of reuse between the original and redesigned projects (\soot vs. \sootup).
% \noindent\textbf{Study Methodology}
Figure~\ref{fig:bi-overview} provides an overview of our study.
We collected redesign-related data from the GitHub repositories of both the original project \soot and the redesigned project \sootup, since developers maintain both projects in parallel during the migration of core functionalities.
We combined keyword and label searches to identify redesign changes scattered around various software artifacts, including source code, pull requests, commits, issues, and documentation. 
For the direction from \sootup to \soot, we collected \soot-related data from the \sootup repository using keywords ``old soot'' and ``soot equivalent''.
Conversely, for the opposite direction, we used keywords ``sootup'' and ``future soot''.
We selected these keywords based on terms used in the projects' own design documentation (e.g., \texttt{index.md} and \texttt{whatsnew.md}).
% We selected these keywords as they have been consistently used by developers when referencing the project.
We also collected all closed issues (i.e., labeled as bugs and unlabeled) to identify residual bugs inherited from \soot and potential opportunities for synchronizing fixes between the two projects.
Two authors performed independent manual analyses and discussed any conflicts to reach a consensus.

\begin{tcolorbox}[left=0pt,right=0pt,top=0pt,bottom=0pt]
\textbf{Finding 1:} 
In the \soot/\sootup redesign, related changes are not localized to code but are fragmented across various types of software artifacts (e.g., commits, pull requests, issues, and documentation). This fragmentation poses challenges in tracking the changes between the original and the redesigned projects. Although keyword and label searches can partially recover these links, such approaches are insufficient at scale, motivating the need for automated tools that explicitly track cross-project change relationships.
\end{tcolorbox}

\subsection{Analysis from \sootup to \soot}
\label{subsec:sootup-to-soot}

\noindent\textbf{Keyword Search:} We collected \numcodeks code instances, \numprks PRs, \numcmtks commits, \numisks issues and \numdocks documents, covering a five-year span of historical data obtained through keyword search.

\noindent\textbf{Label Search:} 
We examined 256 closed issues, labeled as bugs or unlabeled. Among these, \stlabeltotal were referenced by commits (\stlabelbug labeled as bugs, \stnolabel unlabeled).
No reusable patch or test was found, mainly because: (1) keyword matching fails to locate the corresponding file or code in \soot to apply the patch in \sootup; (2) the \sootup bug cannot be replicated in \soot.

\subsection{Analysis from \soot to \sootup}
\noindent\textbf{Keyword Search:} 
The keyword search found two PRs, two commits, three issues, and two documentation entries, but no code instances.
Only a closed PR~\citep{soot1834} addressing the always-false condition (Section~\ref{subsec:sootup-to-soot}) was related to code reuse.
To identify potential test reuse, we matched code and test classes by file name
(\eg  ``JNopStmt.java'' $\rightarrow$ ``JNopStmtTest.java''), revealing 35 test classes missing in \soot \sversion but present in \sootup \stversion.
These can be potentially backported as the corresponding code classes exist in both versions.
% Due to the significant efforts required for manual analysis, we plan to backport these tests in the future. 

\noindent\textbf{Label Search:}
We compiled 678 closed issues over 12 years, categorized as bugs or left unlabeled. Among these, 75 issues were referenced by commits (37 bugs, 38 unlabeled).
We did not identify portable code for \sootup due to (1) non-reproducible bugs, and (2) fixes already existing in \sootup.

We manually analyzed the constructed dataset and observed several characteristics (\textbf{C}) and patterns (\textbf{P}) of software reuse. 
For each characteristic, we took corresponding the action (\textbf{Action}) with carefully reviewed contributions, and reflected upon the results (\textbf{Reflection}). 
In the remaining examples in this paper, red lines denote \soot code, whereas green lines represent \sootup code.

\noindent\textbf{[C1] TODO comments are a hidden form of deferred code reuse
which may include important validation checks.} 
Among the \numcodeks code instances, \numcpcode contain TODO comments 
that embed copied code fragments from Soot as placeholders for future adaptation. These TODOs correspond to  validation methods, which perform integrity and correctness checks on generated Jimple or other intermediate representation (IR) bodies. Although such validation logic constitutes non-functional requirements, it is critical for ensuring the quality of generated IRs. 
However, our study revealed \textbf{\emph{\sootup developers tend to defer 
implementing non-functional requirements (e.g., validation methods)}}, incurring technical debt that may compromise the quality of generated IRs. \underline{\emph{\textbf{Action:}}} To reduce this debt, 
we submitted five PRs to port several validation methods from \soot, including test cases; three have been merged~\citep{sootupidentityvalidator,sootupvalidator866,sootupvalidator872}. As of the release of \sootup \stversion, eight validation methods were adapted, of which six are Type-3 clones and two are Type-4 clones of their \soot counterparts.
\underline{\emph{\textbf{Reflection:}}} Our analysis reveals that \textbf{\emph{deferred code reuse during redesign often goes beyond copy-paste, requiring semantic reinterpretation}} to fit the new context. In practice, \sootup developers adapted code in the TODO comments by
(1) changing method signatures (parameter and return types); 
(2) adapting names of variables and invoked methods (e.g., ``unit'' to ``stmt''); 
(3) adjusting internal API usage (e.g., \soot uses \texttt{body.getMethod()}, while \sootup uses \texttt{view.getMethod(body.getMethodSignature())}); and 
(4) introducing functional divergence, which may result in Type-4 clones (e.g., \soot's MethodDeclarationValidator checks valid method signatures, while \sootup additionally enforces that signatures avoid impossible modifier combinations). 
These changes reduce textual similarity to the original code, making detection more challenging.

\begin{figure}[!ht]
\centering
\begin{lstlisting}[numbers=left, linewidth=\columnwidth, breaklines=true,language=Java,style=mystyle]
public List<ValidationException> validate(Body body,View view){
    // TODO: check copied code from old soot
    /* SootMethod methodRef = body.getMethod(); ... } */
    return null; }
\end{lstlisting}

% \vspace*{-0.2in}
\caption{\sootup method body copied from \soot code as TODO comments for future reuse}
\label{fig:validator}
% \vspace*{-6pt}
\end{figure}

In issues, commits, and PRs, we focused on keyword-related descriptions and patches.
We also obtained all commits related to code reuse that belong to the collected PRs.
Comparing the code in PR patches with the corresponding code in the latest \soot release, we manually identified code clones in 8 PRs.
We excluded 19 PRs as they were either irrelevant to code in \soot or contained enormous changes (i.e., difficult to isolate and pinpoint related changes).  
Within the 8 PRs, we identified the characteristic \textbf{[C2]}.

\noindent\textbf{[C2] Forward porting commonly occurred in the source code of redesigned projects instead of test code, whereas backporting of code/tests rarely occurred.} 
Inspection of PR patches shows that most porting from \soot to \sootup occurred in source code instead of test code, aligning with previous findings that forked projects rarely benefit from tests created in other forks~\citep{mukelabai2023share}. The substantial amount of untested functionalities in \soot highlights \textbf{\textit{missed opportunities for backporting tests}}.
Several factors may limit test reuse: (1) \soot had many commits without tests, indicating inadequate test coverage; (2) significant API changes in \sootup compared to \soot~\citep{karakayasootup} reduced code similarity, limiting test reuse; (3) \sootup meticulously designed its testing framework and test cases.
For example, when porting the \texttt{DeadAssignmentEliminator} class, the \texttt{internalTransform} method in \soot was renamed to \texttt{interceptBody} in \sootup. However, the corresponding test class \texttt{DeadAssignmentEliminatorTest} had completely different tests.
In \soot, each test constructs its own input, whereas \sootup uses fewer inputs but more reusable inputs (\eg several tests share a \texttt{createBody()} helper method~\citep{sootup315}). 
\underline{\emph{\textbf{Action:}}} To encourage backporting, we contributed by backporting bug-fix-related pull requests.
For example, we noticed that a PR was previously submitted to \soot but was closed due to missing tests, while \sootup reused the fix and added corresponding tests. To address this, we submitted a PR to backport both the fix and its tests, which was accepted \citep{sootportfix}. \underline{\emph{\textbf{Reflection:}}} Our experience of backporting reveals that \noindent\textbf{backporting fixes with associated tests may help unblock issues/PRs that were previously closed due to missing tests}, 
which can subsequently help improve the quality of the original project (\soot). 

\begin{figure}[!ht]
\centering
\noindent
\begin{minipage}[t]{0.6\textwidth} % 左侧宽度
\begin{adjustbox}{width=\textwidth, valign=t} % cancel left margin of subfig/minipage, adjust fontsize
\begin{lstlisting}[numbers=left, breaklines=true, language=diff]
private void convert() {     ...
- Edge edge = worklist.pollLast();
- AbstractInsnNode insn = edge.insn;
+ BranchedInsnInfo edge =worklist.pollLast();
+ AbstractInsnNode insn = edge.getInsn();
\end{lstlisting}
\end{adjustbox}
\caption{\footnotesize Identifier name reuse with type change in \sootup~\citep{sootupcommit} compared to \soot}
\label{fig:sim-identifier}
% \captionof{figure}{Identifier name reuse with type change (Edge to BranchedInsnInfo) in \sootup~\citep{sootupcommit} compared to \soot}
\end{minipage}
\hspace{0.05\textwidth} % 两个minipage之间不能加空行, 会被认为段落结束，插入段落间距，导致第二个minipage被推到下一行
\noindent
\begin{minipage}[t]{\textwidth}
\begin{adjustbox}{width=\textwidth, valign=t}
\begin{lstlisting}[numbers=left, breaklines=true, basicstyle=\small\ttfamily, language=diff]
-protected void internalTransform(Body b, String phaseName, Map<String, String> options) ...
+public void interceptBody(@Nonnull Body.BodyBuilder builder) ...
// Make a first pass through the statements, noting the statements we must absolutely keep.
       ... // Stmt is of the form a = a which is useless  ...
       // Remove the dead statements
-      units.retainAll(essential);
+      for (Stmt stmt : stmts) {         ...
\end{lstlisting}
\end{adjustbox}
\caption{\footnotesize Similar inline comments between \sootup~\citep{sootup315} and \soot}
\label{fig:align-comments}
\end{minipage}
% \caption{跨两栏并排示例}
\end{figure}
\vspace{-1em}

\begin{figure}[!ht]
% \begin{adjustbox}{width=0.8\textwidth, valign=t}
% \begin{lstlisting}[linewidth=\columnwidth, breaklines=true, language=java, style=mystyle,numbers=left, xleftmargin=1em, frame=none]
\begin{lstlisting}[
    numbers=left, 
    linewidth=\textwidth, 
    breaklines=true, 
    language=diff, 
    basicstyle=\small\ttfamily, 
    aboveskip=0pt,
    belowskip=0pt,
    lineskip=-1pt,
    xleftmargin=2em
]
/** Utility methods for string manipulations commonly used in Soot. */
 public class StringTools {
- /**Returns fromString, but with non-isalpha() characters printed
-   as <code>'\\unnnn'</code>. 
- Used by SootClass to generate output. */
+ /**Returns fromString, but with non-isalpha() characters printed
+   as <code>'\\unnnn'</code>.*/
    public static String getEscapedStringOf(String fromString) {...}   ...}
\end{lstlisting}
% \end{adjustbox}
% \vspace*{-0.2in}
\caption{Similar Javadoc comments in a \sootup{} patch~\citep{sootup311} and the corresponding in \soot{} (\sversion{})}
\label{fig:sim-javadoc}
% \vspace*{-6pt}
\end{figure}

\noindent\textbf{[C3] Propagation of residual bugs during code migration and detection of new bugs in reused code blocks.}
Similar to the observation in a prior study~\citep{mondal2019empirical}, our analysis revealed that most code clones in \sootup (with respect to \soot) are Type-3 clones, leading to residual bug propagation.
Specifically, we discovered two fixes in \sootup for residual bugs that were initially reported in \soot~\citep{sootupalwaysfalse,sootoldfix}.
As \soot had left these bugs unfixed, they were inherited by \sootup during code migration.
One bug involved an always-false condition in \texttt{AsmMethodSource}, leading to an unreachable branch.
The \sootup PR~\citep{sootupalwaysfalse} referenced a closed \soot PR~\citep{soot1834}, and its fix was a refined version of that \soot PR, with fewer changes and additional tests.
The other bug resulted in a crash during method body retrieval.
\underline{\emph{\textbf{Action:}}} To resolve the identified residual bugs in \soot, we created two PRs to backport the fixes, along with the associated tests, from \sootup. Both PRs were merged by a \soot developer~\citep{sootportalwaysfalse,sootportfix} with positive comments such as \emph{``Thank you for this effort, much appreciated...''}. 
The test for the second residual bug was ignored in \sootup. While trying to forward-port the test, we found a new bug within a reused code block.
The bug was caused by the initialization of a field \texttt{trapHandler} in \soot which involved two steps.
However, in \sootup, a method call \texttt{buildTraps()} was inserted, referencing the field before its initialization, resulting in a null handler exception. We reported this bug~\citep{sootupfixsuggestions}, and a \sootup developer responded quickly by fixing the issue. \underline{\emph{\textbf{Reflection:}}} 
Fixing residual bugs shows that code migration can propagate existing defects and introduce new ones, highlighting the need for \emph{careful review of modifications to reused code to preserve functionality}.

\begin{tcolorbox}[left=0pt,right=0pt,top=0pt,bottom=0pt]
\textbf{Finding 2:} 
Our study reveals three key insights into code reuse in the ongoing \soot/\sootup redesign:
(1) Redesign is not always a linear migration, as parallel maintenance of two projects necessitates bidirectional reuse through forward porting and backporting.
(2) Redesign mainly focuses on forward porting of production code, while test reuse remains limited despite being theoretically easier, likely due to significant changes in \sootup's internal APIs and testing framework that hinder direct test reuse.
(3) Backporting rarely occurred for both production code and tests. Our study reveals that many \soot classes remain untested despite the existence of corresponding tests in \sootup, indicating missed  backporting opportunities that could catch residual bugs and unblock previously closed PRs due to missing tests. 
\end{tcolorbox}

\noindent\textbf{[P1] Customized renaming rules during redesign.} Using keyword search, we observed that many identifiers in \sootup were systematically renamed during redesign, introducing textual divergence. \emph{These renaming patterns break the direct correspondence between text and semantics, highlighting a common source of semantic-textual mismatch during redesign.}
\begin{itemize}[noitemsep,leftmargin=*]
\item To ensure immutability, \sootup uses withers instead of setters  (\eg \texttt{withName} in \sootup maps to \texttt{setName} in \soot)~\citep{karakayasootup}. 

\item Concept replacement (\eg \texttt{View} in \sootup maps to the singleton \texttt{Scene} in \soot, \texttt{BodyInterceptors} replaces the concept of \texttt{BodyTransformer}, and \texttt{Stmt} corresponds to \texttt{Unit}. Consequently, methods like \texttt{getStmt} correspond to \texttt{getUnit}).
\end{itemize}

Based on the matched code, we derived code reuse patterns, illustrated below with examples.

\noindent\textbf{[P2] Similar identifier names and method signatures (8/8).} We noticed that code in \sootup often uses identical or similar identifier names (\eg class/field/method/variable names) and method signatures from \soot. For example, Figure \ref{fig:sim-identifier} shows a code reuse instance in \texttt{AsmMethodSource::convert}, wherein \sootup retains the class name and method signature without alteration. The \emph{names of the local variables} (\eg \texttt{edge}, \texttt{ins}) \emph{remain unchanged despite changes in their data types}.
This likely occurs because during redesign, developers tend to restructure the code by adding new classes while the \emph{underlying natural language channel} remains unchanged (i.e., the variables involved remain the same).

\noindent \textbf{[P3] Similar inline and Javadoc comments (5/8).} 
\sootup developers often reuse comments within methods.
Figure \ref{fig:align-comments} shows matching inline comments despite signature and logic changes, aiding code reuse identification.
Figure~\ref{fig:sim-javadoc} further shows identical class Javadoc but shortened method Javadoc in \sootup, reflecting behavioral or usage changes.

Based on these patterns, we derived rules to improve clone detection in redesigned projects (Table~\ref{tab:pattern-rule}). \textit{R1} applies regex-based identifier renaming ( Table~\ref{tab:custom-rules}). \textit{R2} compares method headers (return type, method name and parameters) and local variables (variable type and name) to identify similar methods as a lightweight alternative to full-body analysis. \textit{R3 and R4} measure documentation and comment similarities, respectively, to capture reused descriptions.

\begin{table}[!t]
\centering
\caption{Rules derived from patterns discovered in our study}
% \vspace*{-6pt}
\label{tab:pattern-rule}

\begin{tblr}{
    width = \columnwidth,
    colspec = {l l X},
    cells = {font=\scriptsize},
    colsep = 2pt, % 调整列间距，列宽=文字内容的宽度+2*colsep
    rowsep = 2pt, 
    % --- merge ---
    cell{4}{1} = {r=2}{m},
    % --- color ---
    row{even} = {bg=gray!10},
    row{1} = {font=\bfseries\scriptsize, bg=gray!30},% header row
    % % --- 线条控制 ---
    hline{1, 2, 6} = {1.2pt, solid}, % 顶线
}
    Pattern & Rule  &  Description  \\ 
    P1 & R1: Redesign-aware Renaming & Renaming identifiers across scopes based on custom rules in redesign documentation. \\ 
    P2 & R2: Similar Identifier Name Matching & Measure similarity in method headers and local variables to identify similar methods. \\ 
    P3 & R3: API Documentation-aware Matching & Measure method API documentation similarity to identify semantically similar methods. \\
       & R4: Inline Comment-aware Matching & Measure inline comment similarities to identify method clones from copy-paste instances. \\
\end{tblr}
\end{table}

\begin{tcolorbox}[left=0pt,right=0pt,top=0pt,bottom=0pt]
\textbf{Finding 3:}
Code reuse in redesign adapts to the new context, often lowering textual similarity to the original code. We identified three textual patterns that preserve semantics and can enhance  clone detection tools for redesigned projects:
(1) customized renaming rules during redesign (\eg using withers instead of setters),
(2) similar identifiers and method signatures despite distinct data types,
and (3) similar comments in code clones across projects.
\end{tcolorbox}

\section{Methodology of Redesign-aware Clone Filtering}
\label{sec:methodology}
\begin{figure}[!hbt]
    \centering
    % \includesvg[width=\linewidth]{fig-ov-approach} % 0.6\linewidth
    \includegraphics[width=\linewidth]{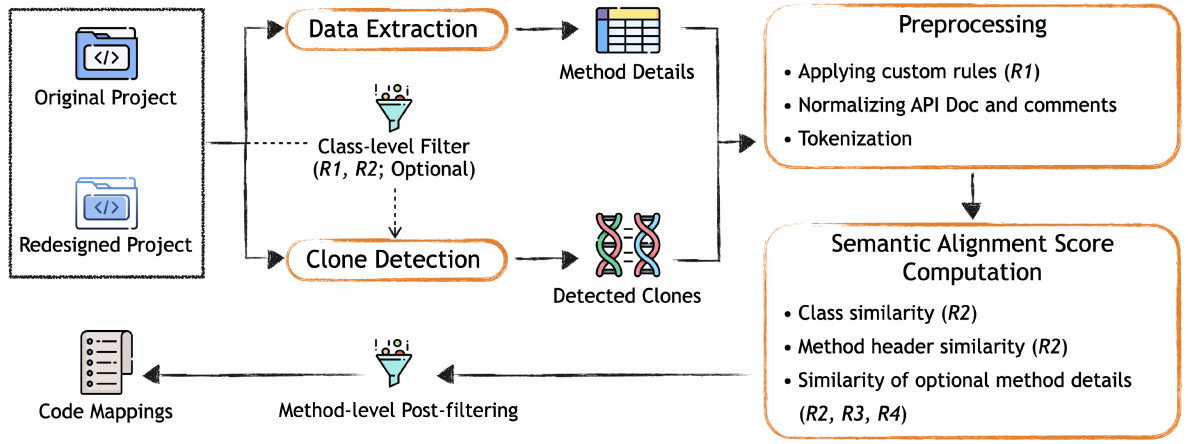} 
    \caption{Overview of our redesign-aware clone filtering workflow that retrofits existing clone detectors.} 
\label{fig:ov-approach}
\end{figure}

Building upon our case study, we extend prior clone detectors with the reuse rules in Table~\ref{tab:pattern-rule} to improve code mapping for redesigned projects.  Here, semantic alignment refers to functional correspondence between method pairs; SAH does not analyze program semantics directly, but uses textual signals as proxies for such correspondence. Figure~\ref{fig:ov-approach} shows our conditional hierarchical workflow: for a redesigned project pair, we extract method details while optionally applying class-level pre-filtering for LLM detectors. We then run clone detectors to identify cross-project method clones. For each candidate pair, we preprocess their details and compute a \textit{semantic alignment score (SAS)}. These scores are then used in a method-level post-filtering step to produce the final code mappings.

\noindent\textbf{\emph{Data Extraction.}}
The data extraction step parses code files to collect method details for similarity computation, including the containing file and class of each method, method header (return type, name, parameters), local variables, inline comments, and docstrings.  We exclude interfaces, abstract methods and constructors, as they lack concrete logic.  We also exclude common methods resulting from object-oriented design (\eg \texttt{toString()} in the \texttt{Object} class) to focus on code mappings between redesigned projects.

\subsection{Class-level Pre-Filtering to Enable Scalable LLM-based Detection}
\label{subsubsec:prefiltering}
To mitigate the pairwise comparison bottleneck of LLM-based clone detectors when scaling to repository-level inputs (CH3), we apply class-level filters. Our goal is to drastically reduce the number of method pairs before feeding them to clone detectors.  Following pattern \textit{P2}, our rationale is that methods belonging to class pairs with low name similarity are unlikely to be correct mappings. We first normalize equivalent concepts in fully qualified class names (\textit{R1}), then compute class name similarity (\textit{R2}) and discard pairs with low scores. 
We only pair methods within each retained class pair, and further discard pairs with a line-count ratio $\ge 2$ or embedding similarity $< 0.5$.

\subsection{Preprocessing}

\noindent\textbf{Applying custom renaming rules \textit{(R1)}:} We identify project-specific renaming conventions in redesign documentation to eliminate irrelevant identifier differences and improve the accuracy of the identifier name matching \textit{(R2)}. 
Table~\ref{tab:custom-rules} lists the renaming rules for the selected project pairs  with the application scope (\eg all details or only method names), target project, and regex patterns for matching and replacement.
For example, we rename \soot setters to \sootup withers via Python regex: \verb|re.sub(r'\bset([A-Z]\w*)', r'with\1', methodName)|.

\begin{table}[!t]
    \centering
    \caption{Implemented redesign-aware replacement rules}
    \label{tab:custom-rules}
    \begin{tblr}{
        width = \columnwidth,
        colspec = {Q[l,wd=3.5em] llll},
        cells = {font=\scriptsize, m},
        colsep = 2pt, % 调整列间距，列宽=文字内容的宽度+2*colsep
        rowsep = 2pt, 
        row{1} = {font=\bfseries\scriptsize, bg=gray!30},% header row
        hline{1, 2, 11} = {1.2pt, solid}, % 顶线
        hline{7,9} = {solid},
        row{odd} = { bg=gray!10},
    }

Scope & Target & Pattern Matching & Replacement & Examples \\ 

Method Name & \soot & \textbackslash{}bset([A-Z]\textbackslash{}w*) & with\textbackslash{}1 & setName $\rightarrow$ withName              \\
All & \soot & (Unit\textbar{}Use\textbar{}Value\textbar{}Def)Box(?:e(?=s))? & \textbackslash{}1 & UnitBox/UnitBoxes $\rightarrow$ Unit/Units  \\
All & \soot & Unit & Stmt & Unit/Units $\rightarrow$ Stmt/Stmts         \\
All & \soot & BodyTransformer & BodyInterceptor & --- \\
All & \sootup & BasicBlock & Block & --- \\

All & \sbugs & \textbackslash{}bConst\textbackslash{}b & Constants & Const.GOTO $\rightarrow$ Constants.GOTO \\
All & \sbugs & spotbugsTestCases & findbugsTestCases & --- \\

All & \alrt & TestASTConstruction & TestASTStructure & --- \\
All & \alrf & TestTokenTypeAssignment & TestSymbolDefinitions & --- \\

    \end{tblr}
\end{table}

% \resizebox{\textwidth}{!}{
% \begin{tabular}{lllll} 
% \hline
% Scope       & Target                 & Pattern Matching                                                                       & Replacement                           & Examples                        \\ 
% \hline
% All         & \soot   & (Unit\textbar{}Use\textbar{}Value\textbar{}Def)Box(?:e(?=s))?                 & \textbackslash{}1     & UnitBox/UnitBoxes $\rightarrow$ Unit/Units  \\
% All         & \soot   & Unit                                                                          & Stmt                                  & Unit/Units $\rightarrow$ Stmt/Stmts         \\
% All         & \soot   & BodyTransformer                                                               & BodyInterceptor                       & -                               \\
% All         & \sootup & BasicBlock                                                                    & Block                                 & -                               \\
% Method Name & \soot   & \textbackslash{}bset([A-Z]\textbackslash{}w*) & with\textbackslash{}1 & setName $\rightarrow$ withName              \\
% \hline
% All         & \sbugs  & \textbackslash{}bConst\textbackslash{}b  & Constants & Const.GOTO $\rightarrow$ Constants.GOTO \\
% All         & \sbugs    & spotbugsTestCases     & findbugsTestCases & - \\
% \hline
% \end{tabular}
% }

\noindent\textbf{Normalizing API docstrings and comments:} API docstrings and inline comments may contain content that is irrelevant to code functionality, so we normalize them before applying the matching rules. For docstrings, we remove inline tags and HTML syntax, comparing only the description text (e.g., \texttt{\{@link Path\}} becomes \texttt{Path}). We also remove URLs and TODOs from docstrings and comments. Moreover, we convert abbreviations in natural language to their full forms to eliminate irrelevant differences (\eg ``\texttt{doesn't}'' is converted to ``\texttt{does not}''). 

\noindent\textbf{Tokenization:} We tokenize method details for similarity measurement by replacing punctuation with whitespace and splitting the text into words. We then split camel case words into individual tokens and convert all tokens to lowercase. 

\subsection{Method-level Post-Filtering}
\label{subsec:post-filtering}

To ensure high-confidence mappings, we apply method-level post-filtering using a \emph{Semantic Alignment Score (SAS)}, which quantifies semantic similarity between method pairs and allows us to filter out candidates below a predefined threshold $thres_{SAS}$. 

\noindent\textbf{Semantic Alignment Score Computation:} SAS quantifies semantic similarity between a method pair ($m_1$, $m_2$) from two projects.
It compares class names, method names, API docstrings, and inline comments, and comprises three components: 
\ding{172} \textbf{Class similarity ($simClass$)} based on class names and docstrings;
\ding{173} \textbf{Method header similarity ($simMethodHeader$)} based on method names, return type, parameter names and types;
\ding{174} \textbf{Optional detail similarity ($simOptional$)} based on names and types of local variable, method docstrings, and inline comments.

Each component is computed using the longest common subsequence (LCS) between token sequences, which has been shown effective for comparing code and documentation~\citep{koznov2024calculating,roy2008nicad}: $\text{sim}(S_1, S_2) = \frac{2 \cdot \text{lcs}(S_1, S_2)}{n + m}$,
where $S_1$ and $S_2$ have $n$ and $m$ tokens, respectively, and $lcs(S_1, S_2)$ is the length of their longest common subsequence. We compute LCS-based similarity for all relevant method details ($simClassName$, $simClassDoc$, $simMethodName$, $simReturnType$, $simParam$, $simLocalVar$, $simMethodDoc$, $simComment$) and aggregate them into SAS components:

\begin{equation}\label{eq:component}
\resizebox{0.9\columnwidth}{!}{$
\left\{
\begin{aligned}
simClass &= simClassName + (1 - simClassName) \cdot simClassDoc,\\
simMethodHeader &= \delta \cdot simMethodName + \eta \cdot simReturnType + \phi \cdot simParam,\\
simOptional &= \text{average}(simLocalVar, simMethodDoc, simComment)
\end{aligned}
\right.
$}
\end{equation}

The overall SAS is a weighted sum of these components:
\begin{equation}\label{eq:combine-sim}
\resizebox{0.8\columnwidth}{!}{$
SAS = \alpha \cdot simClass + \beta \cdot simMethodHeader + \theta \cdot simOptional
$}
\end{equation}

Six weights, $\alpha, \beta, \theta$ and $\delta, \eta, \phi$, control the relative contributions of the similarity components in the equations, subject to $\alpha+\beta+\theta=1$ and $\delta+\eta+\phi=1$, leaving four degrees of freedom.

We tune these weights via grid search on the labeled production-code samples of \soot/\sootup (Table~\ref{tab:samples}) with a step size of 0.05; test-code samples are excluded, as they contain no code mappings. The tuning objective is R-precision, \ie the proportion of true code mappings among the top-$R$ ranked method pairs, where $R$ is the number of true code mappings in the samples, rather than a threshold-dependent metric such as F1-score. This reflects a two-stage design: the weights determine how method pairs are ranked by similarity, while the threshold separately determines the decision boundary applied over this ranking. Jointly optimizing both would enlarge the search space and increase the risk of overfitting on our limited labeled samples. We therefore tune the weights and calibrate the threshold independently: the robustness of weight tuning and threshold generalizability are evaluated separately in Sections~\ref{ssubsec:stability} and~\ref{ssubsec:threshold}, respectively. Ties in R-precision are broken deterministically by first favoring the combination with the largest minimum weight and then, if ties remain, by lexicographic order over the weight tuple.

\noindent\textbf{Implementation.} During data extraction, we use JavaParser~\citep{javaparser} to analyze Java files.
In preprocessing, we apply custom rules (Table~\ref{tab:custom-rules}) to standardize equivalent identifiers for the selected project pairs, prioritizing renaming complex identifiers to simpler ones to reduce incorrect replacements in camel-case identifiers.
For SAS computation, we use the Python package \texttt{pandarallel} to parallelize execution.

\section{Evaluation}
\label{sec:eval}

Our evaluation aims to address the following research questions:

\begin{itemize}
  \item \textbf{RQ1:} \textit{To what extent can existing code clone detectors identify code mappings between original and redesigned projects in production and test code?}
  This RQ examines the effectiveness of existing clone detectors in identifying method-level code mappings across redesigned projects and the challenges they face in such scenarios.

  \item \textbf{RQ2:} \textit{How effective is our filtering approach in identifying code mappings, and how robust and generalizable are its weights and thresholds?}
  This RQ evaluates the effectiveness of our filtering approach in distinguishing code mappings from irrelevant clones, and examines the robustness and generalizability of its tuned weights and thresholds.

  \item \textbf{RQ3:} \textit{What is the contribution of each rule to filtering method clones?} 
  This RQ analyzes the contribution of each rule to the overall approach through an ablation study.

  \item \textbf{RQ4:} \textit{How do SAS-enhanced traditional clone detectors compare with LLM detectors using a CM-oriented prompt?} This RQ provides a task-aligned comparison between SAS-enhanced traditional detectors and LLM detectors for code mapping, complementing the GC-oriented LLM comparison in RQ1 and RQ2.

\end{itemize}

\subsection{Experimental Setup}

\subsubsection{Configurations}
\label{subsubsec:config}

\noindent\textbf{Traditional Baselines.} 
We use four traditional clone detectors as baselines: NiCad, DeepSim, CCStokener, and SSCD (Section~\ref{introclone}). Following prior studies~\citep{testclones2020,testprodclones2021}, we adopt a minimum clone size of 5 LOC for all tools. NiCad and CCStokener use default similarity thresholds of 0.7 and 0.6, respectively. For DeepSim, we use the version trained on Google Code Jam data rather than BigCloneBench, which has been shown to be problematic for learning code similarity~\citep{krinke2022bigclonebench}. For SSCD, we set the similarity threshold to 0.95, the code length to 128 tokens, top-N to 10, and apply whitespace removal preprocessing, following the original evaluation~\citep{sscd2022,sscd2024}. For all other tool configurations, we reuse their default values. This design choice avoids task-specific tuning for individual tools and ensures a fair comparison across different clone detectors.

\noindent\textbf{LLM-based Detection.} We use two open-weight models, GPT-OSS-120B and Qwen2.5-Coder-32B-Instruct (hereafter \emph{GPT-OSS} and \emph{Qwen-Coder}). We select open-weight models with strong code understanding capabilities that are practical for local inference. GPT-OSS-120B is reported to approach closed-source models such as o4-mini on coding and reasoning benchmarks~\citep{gptoss}, while Qwen2.5-Coder-32B-Instruct is a code-specialized model with a different scale and training lineage~\citep{qwencoder}. Both models are evaluated zero-shot without fine-tuning on an NVIDIA H100 GPU.  

We use a temperature of 0.3 for both models. For the GC detection, this matches the optimal temperature reported with the prompt from prior work~\citep{llm4ccd}; we use the same temperature for the CM prompt for consistency. Table~\ref{tab:prompts} presents the two prompt variants. The GC prompt follows prior work~\citep{llm4ccd}, whereas the CM prompt is derived from our definition of code mapping (Section~\ref{sec:intro}). GPT-OSS and Qwen-Coder are used in RQ1 and RQ2 with the GC prompt, whereas RQ4 uses GPT-OSS with the CM prompt.

\noindent\textbf{SAS Configuration.} We use the following configuration throughout the experiments: $\alpha=0.05$, $\beta=0.70$, $\theta=0.25$, $\delta=0.70$, $\eta=0.05$, and $\phi=0.25$, with a threshold of $t=0.50$ for GC and $t=0.70$ for CM. The weights are selected based on the stability analysis in Section~\ref{ssubsec:stability}, while the thresholds are derived in Section~\ref{ssubsec:threshold}.

\begin{table}[ht]
    \centering
    \small
    \caption{Prompt variants used by the LLM detector.}
    \label{tab:prompts}
    \begin{tblr}{
        width = \columnwidth,
        colspec = {l X},
        cells = {font=\footnotesize,m},
        hlines,
        row{1} = {font=\bfseries\footnotesize, bg=gray!30}
    }

    Prompt Variant & Prompt \\
    GC prompt & 
        \ttfamily Do Code1 and Code2 solve identical problems with the same inputs and outputs? Answer ONLY with ``yes'' or ``no''. Do NOT write anything else. No explanations. \\
    CM prompt & 
        \ttfamily Method 1 is from a redesigned version of a software project. Method 2 is from the project's original version. Redesign may change data structures, types, method signatures, or implementation details while preserving the underlying purpose of the code. Does Method 1 represent the functional migration of Method 2 (same underlying purpose, even if inputs/outputs differ due to redesign)? Answer ONLY with ``yes'' or ``no''. Do NOT write anything else. No explanations. \\
    
    \end{tblr}
\end{table}

\subsubsection{Dataset Construction}
\label{subsec:benchmark}

\begin{table}[!t]
    \centering
    \begin{talltblr}[
        theme = mockcaption,
        caption = {Labeling results for method pairs in sample set $D$.},
        label = {tab:samples},
        remark{Notation}={T1--T4 denote clone types (Types 1--4); T1 (\%)--T4 (\%) denote the percentage of each type that are code mappings.},
    ]{
        width = \columnwidth,
        colspec = {l l r X[r] r r r r r r r r r r},
        cells = {font=\scriptsize},
        colsep = 2pt, % 调整列间距，列宽=文字内容的宽度+2*colsep
        rowsep = 2pt, 
        % --- merge ---
        cell{1}{1,2,3,4} = {r=2}{m},% code type, pair, total
        cell{1}{5,10} = {c=5}{c},% genuine clones, code mappings
        cell{3}{1} = {r=4}{m},% prod
        cell{7}{1} = {r=4}{m},% test
        % --- color ---
        row{1,2} = {halign=c, font=\bfseries\scriptsize, bg=gray!30},% header row
        cell{6,10}{2} = {font=\bfseries\scriptsize}, % all pairs
        row{6,10} = {bg=gray!10},
        % % --- 线条控制 ---
        hline{1, 3, 7, 11} = {1.2pt, solid}, % 顶线
        hline{2,6,10} = {solid}, % 
        vline{5,10} = {solid},
    }

Type & Pair & Total & NonClone & Genuine Clone & & & & & Code Mapping & & & & \\
 & & & & Total & T1 & T2 & T3 & T4 & Total & T1(\%) & T2(\%) & T3(\%) & T4(\%) \\

Prod & \soot/\sootup & 400 & 244 & 156 & 41 & 42 & 39 & 34 & 127 & 100.00 & 57.14 & 84.62 & 85.29 \\
Prod & \fbugs/\sbugs & 400 & 124 & 276 & 171 & 21 & 81 & 3 & 247 & 100.00 & 47.62 & 77.78 & 100.00 \\
Prod & \alrt/\alrf & 244 & 197 & 47 & 6 & 0 & 29 & 12 & 30 & 100.00 & 0.00 & 51.72 & 75.00 \\
Prod & All Pairs & 1044 & 565 & 479 & 218 & 63 & 149 & 49 & 404 & 100.00 & 53.97 & 74.50 & 83.67 \\
Test & \soot/\sootup & 168 & 168 & 0 & 0 & 0 & 0 & 0 & 0 & 0.00 & 0.00 & 0.00 & 0.00 \\
Test & \fbugs/\sbugs & 324 & 103 & 221 & 57 & 28 & 132 & 4 & 118 & 100.00 & 82.14 & 25.76 & 100.00 \\
Test & \alrt/\alrf & 269 & 196 & 73 & 13 & 31 & 24 & 5 & 53 & 92.31 & 83.87 & 41.67 & 100.00 \\
Test & All Pairs & 761 & 467 & 294 & 70 & 59 & 156 & 9 & 171 & 98.57 & 83.05 & 28.21 & 100.00 \\

% Prod & \soot/\sootup & 400 & 258 & 142 & 37 & 41 & 35 & 29 & 114 & 100.00 & 56.10 & 82.86 & 86.21 \\
% Prod & \fbugs/\sbugs & 535 & 194 & 341 & 206 & 34 & 95 & 6 & 299 & 100.00 & 44.12 & 78.95 & 50.00 \\
% Prod & \alrt/\alrf & 218 & 174 & 44 & 6 & 0 & 28 & 10 & 27 & 100.00 & 0.00 & 50.00 & 70.00 \\
% Prod & All Pairs & 1153 & 626 & 527 & 249 & 75 & 158 & 45 & 440 & 100.00 & 50.67 & 74.68 & 77.78 \\
% Test & \soot/\sootup & 107 & 107 & 0 & 0 & 0 & 0 & 0 & 0 & 0.00 & 0.00 & 0.00 & 0.00 \\
% Test & \fbugs/\sbugs & 466 & 245 & 221 & 57 & 28 & 132 & 4 & 118 & 100.00 & 82.14 & 25.76 & 100.00 \\
% Test & \alrt/\alrf & 292 & 220 & 72 & 13 & 31 & 23 & 5 & 53 & 92.31 & 83.87 & 43.48 & 100.00 \\
% Test & All Pairs & 865 & 572 & 293 & 70 & 59 & 155 & 9 & 171 & 98.57 & 83.05 & 28.39 & 100.00 \\
    \end{talltblr}
\end{table}

\begin{table}[!t]
    \centering
    \begin{talltblr}[
        theme = mockcaption,
        caption = {Inter-annotator agreement across the six data subsets.}, 
        label = {tab:kappa},
        remark{Note}={Numbers in parentheses indicate disagreement counts for GC/CM and agreement counts for Exact Agreement. ``--'' denotes undefined $\kappa$ for $d_4$, where all annotations fall into a single category.},
    ]{
        width = \columnwidth,
        % colspec = {X[l, 4] *{6}{X[r]}}, % 使用 X 列撑满指定宽度
        colspec = {X[l] *{6}{r}},
        cells = {m},
        colsep = 2.2pt, % 调整列间距，列宽=文字内容的宽度+2*colsep
        rowsep = 2pt, 
        % % --- color ---
        row{1} = {bg=gray!30},% header row
        % % % --- 线条控制 ---
        hline{1, 2, 5} = {1.2pt, solid}, % 顶线
        hline{4} = {solid},
    }

             & $d_1$ & $d_2$ & $d_3$ & $d_4$ & $d_5$ & $d_6$ \\

GC $\kappa$ & 0.76\textcolor{gray}{(65)} & 0.40\textcolor{gray}{(147)} & 0.88\textcolor{gray}{(28)} & -- & 0.85\textcolor{gray}{(13)} & 0.84\textcolor{gray}{(21)} \\
CM $\kappa$ & 0.88\textcolor{gray}{(22)} & 0.96\textcolor{gray}{(6)} & 0.93\textcolor{gray}{(12)} & -- & 0.94\textcolor{gray}{(3)} & 0.87\textcolor{gray}{(11)} \\
Exact Agreement (\%) & 82.25\textcolor{gray}{(329)} & 54.01\textcolor{gray}{(175)} & 91.75\textcolor{gray}{(367)} & 100.00\textcolor{gray}{(168)} & 93.85\textcolor{gray}{(229)} & 89.22\textcolor{gray}{(240)} \\

    \end{talltblr}
\end{table}

We evaluated baseline tools and our approach on three redesign pairs: \soot/\sootup, \fbugs/\sbugs, and \alrt/\alrf.

\noindent\textbf{Ground Truth Dataset.}
As manually examining all clone pairs is impractical, we built a ground truth dataset $D$ through manual inspection, comprising six subsets from production and test code for each redesign pair. The sampled candidates came from four traditional clone detectors and GPT-OSS, representing the LLM-based technique considered in our evaluation; the Qwen-Coder variant was excluded from sampling.
We targeted approximately 400 randomly sampled candidates per subset (80 per baseline), a sample size commonly used to achieve a 95\% confidence level with a 5\% margin of error, though the actual number varies across sets due to differences in available clone candidates (Table~\ref{tab:postfilt-raw}, ``Total'' column).
For each baseline, we first excluded candidates already present in $D$ to avoid duplicate labeling, then randomly sampled from the remaining candidates up to the 80-sample quota. When fewer than 80 candidates were available, we sampled all available candidates without adjustment, so the resulting subset size may fall below 400. As new baselines, redesign pairs, or updates to the candidate-generation process were introduced over the course of the study, newly sampled candidates were labeled and added to $D$, while previously labeled samples were retained to avoid redundant manual effort.
To avoid circularity between rule derivation and evaluation, we excluded methods introduced by our own ported patches and tests, as our heuristic rules were derived from manual analysis of these patches and some of the corresponding PRs have since been merged into the latest \soot/\sootup versions. This process resulted in \numdataset labeled samples in total.
% Two authors independently reviewed all sampled method pairs and met to resolve disagreements to ensure consistent clone labels. 

\noindent\textbf{Annotation Procedure.} The first two authors, both PhD students in software engineering with relevant domain knowledge, independently labeled each sampled method pair for its clone type (non-clone or Type-1--Type-4) and whether it constitutes a code mapping (CM), following the definitions in Sections~\ref{sec:intro} and ~\ref{sec:background}.

For candidate pairs not identified as direct code mappings, we additionally checked for transitive mappings. Given a candidate pair $(m_1, m_2)$, we searched for the corresponding method $m_2'$ in the same project as $m_1$ based on method and file names. The pair was labeled as a CM via transitivity if $m_1$ and $m_2'$ formed a Type-1 within-project clone and $m_2'$ and $m_2$ were an established code mapping; only Type-1 within-project clones triggered transitivity. The same procedure was applied symmetrically in the other direction.

We partitioned $D$ into six subsets, $d_1$--$d_6$, and computed Cohen's $\kappa$~\citep{cohenkappa2010} separately for GC and CM after labeling each subset. Table~\ref{tab:kappa} reports the agreement statistics and disagreement counts before resolution. All disagreements were resolved through discussion. CM agreement was almost perfect across all subsets ($\kappa=0.87$--$0.96$). GC agreement was substantial to almost perfect in four of the five applicable subsets ($\kappa=0.76$--$0.88$) but moderate in $d_2$ ($\kappa=0.40$), the first test-code subset. Test code introduced ambiguities because superficially similar setup and assertion code can target different components, granularities, or validation objectives. We therefore refined the labeling procedure with test-specific criteria requiring correspondence in these aspects and applied them to subsequent test-code subsets. The complete guidelines and annotated examples are provided in the replication package~\citep{replication}.

\subsubsection{Observations from the Dataset}

% \begin{figure}[!ht]
% \centering
%     % edu.umd.cs.findbugs.ba.ch.Subtypes2Test
% \begin{lstlisting}[linewidth=\columnwidth, breaklines=true, language=java, style=mystyle,numbers=left, xleftmargin=1em, frame=none]
% + void testStringSubtypeOfObject() throws Throwable {
% - public void testArrayOfPrimitiveIsSubtypeOfObject() throws Throwable {
%     executeFindBugsTest(new RunnableWithExceptions() {
%       @Override
% -     public void run() throws Exception {
% +     public void run() throws Throwable {
%         Subtypes2 test = getSubtypes2();
% -       assertTrue(test.isSubtype(typeArrayInt, typeObject)); }});}
% +       Assertions.assertTrue(test.isSubtype(typeString, typeObject)); }});}
% \end{lstlisting}
%     \caption{An example of uninteresting clones in test code of \fbugs/\sbugs}
%     \label{fig:coincident-clones}
% % \vspace*{-6pt}
% \end{figure}

\begin{figure*}[!htb]
\centering

\begin{subfigure}[b]{\textwidth}
    \centering
\begin{adjustbox}{width=\textwidth, valign=b}
    % edu.umd.cs.findbugs.ba.ch.Subtypes2Test
\begin{lstlisting}[breaklines=true, language=java, style=mystyle,numbers=left, xleftmargin=1em, frame=none]
- public void testArrayOfPrimitiveIsSubtypeOfObject() throws Throwable {
+ void testStringSubtypeOfObject() throws Throwable {
    executeFindBugsTest(new RunnableWithExceptions() {
      @Override
-     public void run() throws Exception {
+     public void run() throws Throwable {
        Subtypes2 test = getSubtypes2();
- assertTrue(test.isSubtype(typeArrayInt,typeObject));}});}
+ Assertions.assertTrue(test.isSubtype(typeString,typeObject));}});}
\end{lstlisting}
\end{adjustbox}
\end{subfigure}
\hfill    
\begin{subfigure}[b]{0.8\textwidth}
    \centering
    \begin{adjustbox}{width=\textwidth, valign=b}
    \includegraphics[width=\textwidth]{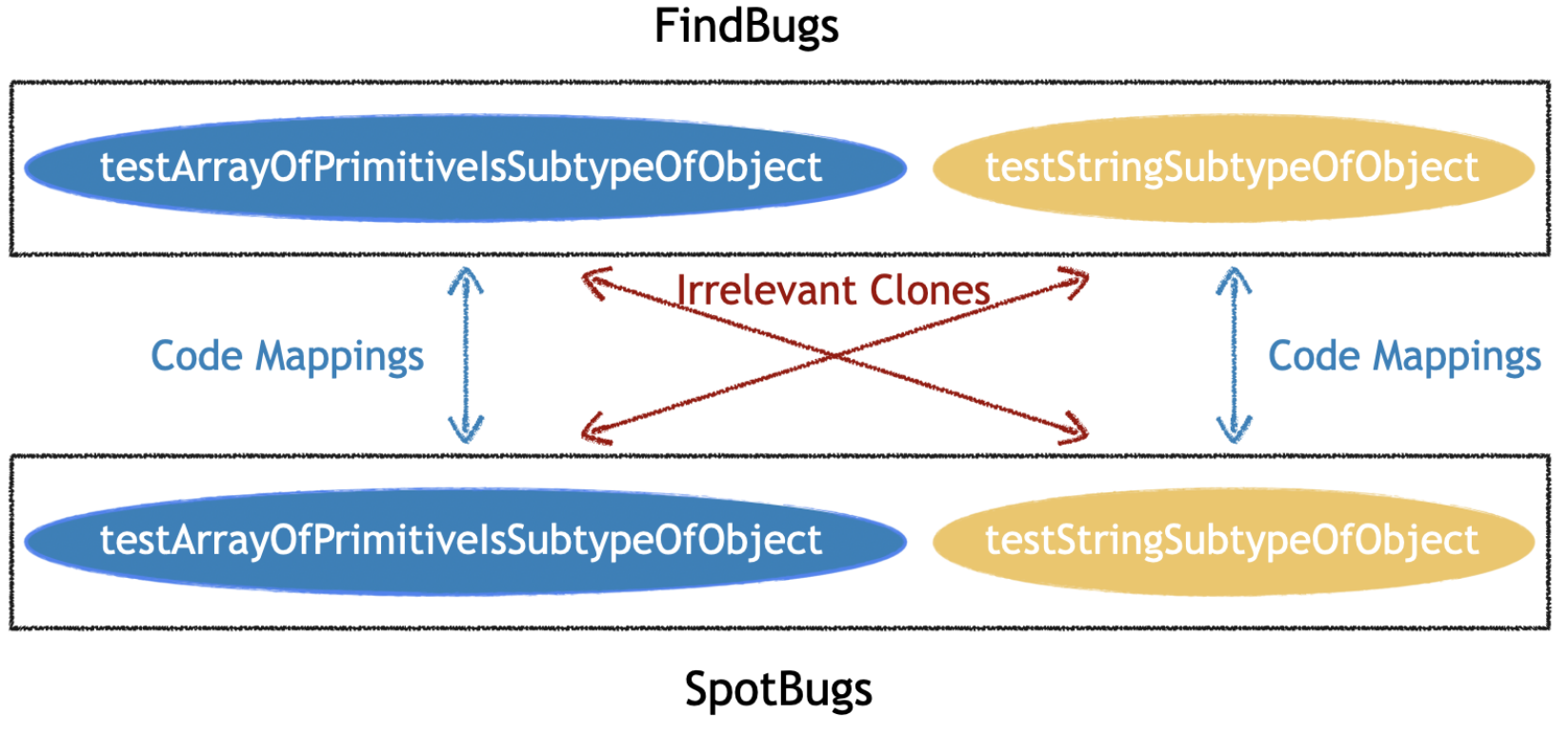}
    \end{adjustbox}
\end{subfigure}
\caption{An example of irrelevant clones that are not code mappings in test code of \fbugs/\sbugs}
\label{fig:test-clones}
\end{figure*}

\begin{figure*}[!htb]
\centering

\begin{subfigure}[b]{0.49\textwidth}
    \centering
    % qilin/core/builder/CallGraphBuilder.java
\begin{lstlisting}[linewidth=\columnwidth, breaklines=true, language=java, style=mystyle,numbers=left, xleftmargin=1em, frame=none]
public OnFlyCallGraph getCICallGraph() {
 if (cicg == null) {
  constructCallGraph(); }
  return cicg; }
\end{lstlisting}
    \caption{\sootup method in CallGraphBuilder.java}
\end{subfigure}
\hfill    
\begin{subfigure}[b]{0.49\textwidth}
    \centering
    % soot/SootMethod.java
\begin{lstlisting}[linewidth=\columnwidth, breaklines=true, language=java, style=mystyle, numbers=left, xleftmargin=1em, frame=none]
public List<SootClass> getExceptions(){
 if (exceptions == null) {
 exceptions=new ArrayList<SootClass>();}
 return exceptions; }
\end{lstlisting}
    \caption{\soot method in SootMethod.java}
\end{subfigure}
% \vspace{-3mm}
\caption{A non-clone pair identified by NiCad due to high syntactic similarity}
\label{fig:not-clone}
\end{figure*}

\begin{figure*}[!htb]
\centering

\begin{subfigure}[b]{0.49\textwidth}
    \centering
    % qilin/core/builder/CallGraphBuilder.java
\begin{lstlisting}[escapechar=@,breaklines=true, numbers=left,frame=none,style=mystyle]
public Type getType() {
 if(isIntLikeType(op2Box.getValue().getType())) {
   final Type t1 = op1Box.getValue().getType();
   if (isIntLikeType(t1)) {
     return IntType.v(); }
   final LongType tyLong = LongType.v();
   if (tyLong.equals(t1)) {
     return LongType.v(); }}
   return UnknownType.v(); }
\end{lstlisting}
    \caption{\soot method in JUshrExpr.java}
    \label{fig:problem-s}
\end{subfigure}
\hfill    
\begin{subfigure}[b]{0.49\textwidth}
    \centering
    % soot/SootMethod.java
\begin{lstlisting}[ escapechar=@,breaklines=true, frame=none,style=mystyle, numbers=left]
public Type getType() {
  Value op1 = getOp1();
  Value op2 = getOp2();
  if (@\colorbox{yellow}{Type.isIntLikeType(op2.getType())}@) {
    return UnknownType.getInstance();  }
  if (Type.isIntLikeType(op1.getType())) {
    return PrimitiveType.getInt();  }
  if (op1.getType().equals(PrimitiveType.getLong())) {
    return PrimitiveType.getLong();  }
  return UnknownType.getInstance(); }
\end{lstlisting}
    \caption{The clone in JUshrExpr.java in \sootup}
    \label{fig:problem-st}
\end{subfigure}
% \vspace{-3mm}
\caption{An example of a bug in correctly mapped clones}% that indicate bugs when porting functionalities}
\label{fig:problem-clones}
\end{figure*}

% don't know why the highlight display at the wrong place
%   @\bh@ if (Type.isIntLikeType(op2.getType())) @\eh@{

\noindent\textbf{Irrelevant Clones and Non-Clones.}
Table~\ref{tab:samples} shows that most GCs in production code are CMs (404/479, 84.34\%), supporting the applicability of clone detectors for identifying CMs in redesigned projects. In contrast, Type-3 clones dominate test code (156/294 GCs across all pairs), but only 28.21\% are CMs, suggesting that many detected similarities arise from common structures in test harnesses. For example, tests in Figure~\ref{fig:test-clones} are flagged as clones by all traditional baselines, but their method and variable names indicate different functionalities. Such irrelevant clones are misleading for code mapping because mappings for both methods already exist.
Our manual analysis also found numerous non-clones, with three notable characteristics: (1) methods with low identifier-name and functional similarity, (2) short methods with high syntactic similarity (e.g., null checks in Figure~\ref{fig:not-clone}), and (3) methods labeled as clones by learning-based tools but substantially different in size and content. These observations motivate filtering both irrelevant clones and non-clones using identifier-name similarity.

\begin{tcolorbox}[left=0pt,right=0pt,top=0pt,bottom=0pt]
\textbf{Finding 4:} 
Test harnesses in test code can introduce substantial irrelevant clones, complicating code mapping detection. Identifier name similarity helps distinguish both irrelevant clones and non-clones.
\end{tcolorbox}

\noindent\textbf{Bugs Detected in Code Mappings.}
While inspecting code mappings, we found three clones~\citep{sootupjushr,sootupnullcheck,sootuptypecheck} with inconsistent behavior, suggesting potential bugs introduced during redesign. Although CCStokener and DeepSim can detect these mappings, they are buried among approximately 2K and 2M reported clones, respectively (Table~\ref{tab:postfilt-raw}), making them difficult to identify manually. Our enhanced rules help prioritize such mappings for inspection.
For example, a \soot method in Figure~\ref{fig:problem-s} performs extra checks for an \texttt{Int}-like type, while its rewritten Type-4 clone in \sootup (Figure~\ref{fig:problem-st}) returns \texttt{UnknownType}. The highlighted condition should be negated. We also found clones missing a null check or an \texttt{instanceof} check. These issues were reported with fixes to \sootup and accepted. These examples show that examining code mappings can reveal bugs introduced during code porting.

\begin{tcolorbox}[left=0pt,right=0pt,top=0pt,bottom=0pt]
\textbf{Finding 5:} Manual inspection of code mappings revealed three instances of inconsistent behavior, showing that code mappings can help reveal porting-related bugs.
\end{tcolorbox}

\subsubsection{Evaluation Metrics}

We evaluate the baselines and our approach on dataset $D$ under two detection objectives: genuine clones (GC) and code mappings (CM). Each objective defines its positive and negative classes: genuine clones and non-clones for the GC objective, and code mappings and non-mappings for the CM objective, respectively. All confusion-matrix entries are defined over $D$: true positives (TP), false positives (FP), true negatives (TN), and false negatives (FN) denote positive instances correctly detected, negative instances incorrectly detected, negative instances correctly ignored, and positive instances missed by a detector, respectively. We evaluate detection using the following metrics.

\noindent \textbf{FPR:} False Positive Rate quantifies the fraction of negative instances incorrectly classified as positive. It is calculated as $FPR = \frac{FP}{FP + TN}$. 

\noindent \textbf{Precision:} Precision measures the \emph{fraction of correctly predicted positive instances among all instances predicted as positive}, and its formula is $Precision = \frac{TP}{TP + FP}$.

\noindent \textbf{Recall:} Recall computes the \emph{fraction of actual positive instances} that are correctly detected. It is calculated as $\frac{TP}{TP + FN}$.

\noindent \textbf{Average F1-Score:} The F1-score denotes the harmonic mean of precision and recall ($F1 = \frac{2 \cdot Precision \cdot Recall}{Precision+ Recall}$) for a target class. It therefore evaluates only the target class and does not account for the other class. To better measure the overall effectiveness, we follow prior work~\citep{avgf1} and compute two F1 variants by alternately treating the positive and negative classes as the target class. The average F1-score is then calculated as the \emph{mean of these two values}: $Average\ F1= \frac{F1_{positive} + F1_{negative}}{2}$.

% \vspace{-1em}
\subsection{RQ1: Efficacy of Existing Clone Detectors}
\label{sec:rq3}

\begin{table}[!t]
\centering
\begin{talltblr}[
    theme = mockcaption,
    caption = {Effectiveness of code clone detectors for the CM objective on the dataset $D$.},
    label = {tab:baselines-cm},
    note{1}={\soot/\allowbreak\allowbreak\sootup test samples contain no code mappings, capping average F1 at 0.5.},
    note{2}={``All Pairs'' aggregates samples across all project pairs without distinguishing individual pairs.},
    remark{Metrics}={Prec., Rec., and Avg. F1 denote precision, recall, and average F1-score, respectively.},
]{
    width = \columnwidth,
    colspec = {X l r r r r r r r r},
    cells = {font=\footnotesize, m},
    colsep = 3pt, % 调整列间距，列宽=文字内容的宽度+2*colsep
    rowsep = 2pt, 
    % % --- Merged Cell ---
    cell{1}{3,7} = {c=4}{m},
    cell{1}{1,2} = {r=2}{m}, % pair, tool
    cell{3}{1} = {r=6}{m}, % soot/sootup
    cell{9}{1} = {r=6}{m},
    cell{15}{1} = {r=6}{m},
    cell{21}{1} = {r=6}{m, font=\bfseries\footnotesize}, % all pairs
    % % % --- color ---
    % row{odd} = {bg=gray!10}, 
    % column{1,2} = {bg=white},
    row{1,2} = {halign=c, font=\bfseries\footnotesize, bg=gray!30},% header row
    row{21,22,23,24,25,26} = {bg=gray!10},
    % % --- 线条控制 ---
    hline{1, 3, 27} = {1.2pt, solid}, % 顶线
    hline{2,9,15,21} = {solid},
    vline{3,7} = {solid},
}

Pair & Tool & Production Code & &     &    & Test Code &      &      &          \\
     &      & FPR & Prec. & Rec. & Avg. F1 & FPR & Prec. & Rec. & Avg. F1  \\

 \soot/\sootup & NiCad & 0.30 & 0.47 & 0.58 & 0.629 & 0.01 & 0.00 & 0.00 & 0.497 \\
 \soot/\sootup & DeepSim & \textbf{0.38} & 0.34 & 0.41 & 0.513 & \textbf{0.56} & 0.00 & 0.00 & 0.306 \\
 \soot/\sootup & CCStokener & 0.33 & 0.44 & 0.57 & 0.605 & 0.04 & 0.00 & 0.00 & 0.491 \\
 \soot/\sootup & SSCD & 0.15 & 0.74 & \textbf{0.87} & \textbf{0.846} & 0.48 & 0.00 & 0.00 & 0.344 \\
 \soot/\sootup & GPT-OSS & 0.11 & 0.78 & 0.79 & 0.839 & 0.00 & 0.00 & 0.00 & \textbf{0.500} \\
 \soot/\sootup & Qwen-Coder & 0.00 & \textbf{0.98} & 0.46 & 0.755 & 0.00 & 0.00 & 0.00 & \textbf{0.500} \\
 \fbugs/\sbugs & NiCad & 0.12 & 0.91 & 0.80 & 0.823 & 0.17 & 0.51 & 0.31 & 0.565 \\
 \fbugs/\sbugs & DeepSim & \textbf{0.67} & 0.55 & 0.50 & 0.416 & 0.45 & 0.19 & 0.19 & 0.367 \\
 \fbugs/\sbugs & CCStokener & 0.41 & 0.77 & 0.85 & 0.731 & \textbf{0.58} & 0.35 & 0.55 & 0.464 \\
 \fbugs/\sbugs & SSCD & 0.23 & 0.88 & \textbf{1.00} & 0.900 & 0.49 & 0.53 & \textbf{0.97} & 0.679 \\
 \fbugs/\sbugs & GPT-OSS & 0.02 & 0.99 & 0.93 & \textbf{0.948} & 0.01 & 0.98 & 0.82 & \textbf{0.920} \\
 \fbugs/\sbugs & Qwen-Coder & 0.00 & \textbf{1.00} & 0.88 & 0.926 & 0.00 & \textbf{1.00} & 0.47 & 0.756 \\
 \alrt/\alrf & NiCad & 0.00 & 0.92 & 0.37 & 0.739 & 0.16 & 0.36 & 0.36 & 0.601 \\
 \alrt/\alrf & DeepSim & \textbf{0.58} & 0.11 & 0.53 & 0.377 & \textbf{0.81} & 0.21 & 0.85 & 0.323 \\
 \alrt/\alrf & CCStokener & 0.46 & 0.19 & 0.77 & 0.497 & 0.58 & 0.27 & \textbf{0.89} & 0.500 \\
 \alrt/\alrf & SSCD & 0.28 & 0.30 & \textbf{0.87} & 0.638 & 0.16 & 0.57 & 0.87 & 0.795 \\
 \alrt/\alrf & GPT-OSS & 0.02 & 0.76 & 0.43 & \textbf{0.753} & 0.03 & 0.85 & 0.77 & \textbf{0.884} \\
 \alrt/\alrf & Qwen-Coder & 0.00 & \textbf{1.00} & 0.27 & 0.686 & 0.01 & \textbf{0.92} & 0.62 & 0.845 \\
 All Pairs & NiCad & 0.16 & 0.73 & 0.70 & 0.771 & 0.12 & 0.44 & 0.32 & 0.610 \\
 All Pairs & DeepSim & \textbf{0.51} & 0.37 & 0.47 & 0.474 & \textbf{0.61} & 0.16 & 0.39 & 0.360 \\
 All Pairs & CCStokener & 0.39 & 0.55 & 0.76 & 0.664 & 0.43 & 0.31 & 0.65 & 0.553 \\
 All Pairs & SSCD & 0.21 & 0.74 & \textbf{0.95} & 0.849 & 0.36 & 0.43 & \textbf{0.94} & 0.679 \\
 All Pairs & GPT-OSS & 0.06 & 0.91 & 0.85 & \textbf{0.901} & 0.02 & 0.94 & 0.81 & \textbf{0.917} \\
 All Pairs & Qwen-Coder & 0.00 & \textbf{1.00} & 0.70 & 0.869 & 0.01 & \textbf{0.97} & 0.52 & 0.805 \\
\end{talltblr}
\end{table}  % 对于同一个 source–target pair，哪个 detector 表现最好?

\noindent\textbf{Approach.} Using the same six clone detectors and configurations described in Section~\ref{subsubsec:config}, we directly evaluate their raw GC-oriented detection outputs for identifying code mappings in the labeled dataset $D$ described in Section~\ref{subsec:benchmark}. This assesses how well existing clone detectors designed to identify genuine clones can be directly applied to the CM task.

\noindent\textbf{Results.}
Table~\ref{tab:baselines-cm} shows that GPT-OSS achieves the highest average F1 at the ``All Pairs'' level (0.917 on production code) among all detectors, combining high precision, low FPR, and the high recall. Qwen-Coder attains comparable precision but a lower average F1 (0.805 on test code), driven by lower recall. Among the traditional detectors, SSCD achieves the highest average F1, driven by its high recall, which is accompanied by a higher FPR. NiCad has the lowest FPR among traditional detectors and a more balanced profile across the four metrics. DeepSim performs substantially worse, with the highest FPR and lowest precision among all detectors on both code types. These results reveal distinct weaknesses across detectors: DeepSim produces excessive false positives; NiCad is relatively conservative, particularly on test code where recall drops to 0.32; CCStokener and SSCD achieve higher recall but exhibit higher FPR; and Qwen-Coder similarly favors precision over recall, most notably on test code.

A consistent pattern also emerges across code types: five of the six detectors perform noticeably worse on test code than on production code, with average F1 dropping by 0.064--0.17 (e.g., SSCD from 0.849 to 0.679). GPT-OSS is the only exception, maintaining essentially the same performance across both code types (0.901 vs. 0.917), suggesting it is comparatively robust to the structural and stylistic differences between production and test code that appear to degrade the other detectors.
Moreover, high clone-detection performance does not necessarily translate into mapping performance. For example, CCStokener achieves 0.63 precision on test code under the GC objective (Table~\ref{tab:baselines-gc}, All Pairs), but only 0.31 under the CM objective, showing that clone similarity alone is insufficient to identify code mappings reliably. These limitations motivate the use of an additional scoring mechanism, as provided by SAS.

\begin{tcolorbox}[left=0pt,right=0pt,top=0pt,bottom=0pt]
\textbf{Finding 6:} Existing clone detectors are more effective at identifying code mappings in production code than in test code. LLM detectors achieve the highest average F1 (0.90/0.92 for GPT-OSS), while the best traditional detector, SSCD, reaches 0.85/0.68. Detector-specific false positives and missed mappings limit the reliability of traditional detectors, motivating adaptation for more effective mapping detection.
\end{tcolorbox}

\subsection{RQ2: Effectiveness and Generalizability of SAS Filtering}
\label{sec:rq2}

\subsubsection{Pre-filtering Effectiveness} 
\label{subsec:prefilt}

\begin{table}[!t]
    \centering
    \begin{talltblr}[
        theme = mockcaption,
        caption = {Number of comparisons remaining after pre-filtering and LLM detector runtime.},
        label = {tab:prefilt},
        % note{1} = {Reported time is wall-clock execution time and may include model loading, I/O, and other overhead beyond inference itself.},
        % note{2} = {GPT-OSS time is averaged over the CM and GC prompts.},
        remark{Note}={Reported time is wall-clock execution time and may include model loading, I/O, and other overhead beyond inference itself. GPT-OSS time is averaged over the CM and GC prompts.},
    ]{
        width = \columnwidth,
        colspec = {l r r r r r},
        cells = {font=\footnotesize},
        % colsep = 2pt, % 调整列间距，列宽=文字内容的宽度+2*colsep
        rowsep = 2pt, 
        % % --- merge ---
        cell{1}{1} = {r=2}{m},% pair
        cell{1}{2} = {c=3}{c},% comparison
        cell{1}{5} = {c=2}{m},% tiem
        % cell{7}{1} = {r=4}{m},% test
        % % --- color ---
        row{1,2} = {halign=c, font=\bfseries\footnotesize, bg=gray!30},% header row
        % % % --- 线条控制 ---
        hline{1, 3, 6} = {1.2pt, solid}, % 顶线
        hline{2} = {solid}, 
        vline{2,5} = {solid},
    }

Pair & Comparisons & & & Time (hours)  &   \\
     & Total  & Kept & Drop (\%) & GPT-OSS & Qwen-Coder \\

\soot/\sootup & 12,258,183 & 112,039 & 99.09 & 13.03 & 3.09 \\
\fbugs/\sbugs & 15,589,480 & 272,846 & 98.25  & 42.76 & 5.98 \\
 \alrt/\alrf & 1,192,818 & 209,253 & 82.46  & 23.33 & 4.29 \\ 
    \end{talltblr}
\end{table}

We applied class-level pre-filtering (Section~\ref{subsubsec:prefiltering}) to reduce the number of method-pair comparisons for LLM detectors. As shown in Table~\ref{tab:prefilt}, pre-filtering eliminates 99\%, 98\%, and 82\% of the comparisons for \soot/\sootup, \fbugs/\sbugs, and \alrt/\alrf, respectively. Despite these substantial reductions, 112K--273K comparisons remain for each project pair, representing a considerable workload for LLM-based detection. This motivates our use of open-weight models that can be deployed locally, avoiding dependence on per-query costs and API-based processing at this scale. With local inference, GPT-OSS processes the remaining comparisons in approximately 13, 43, and 23 hours for \soot/\sootup, \fbugs/\sbugs, and \alrt/\alrf, respectively, while Qwen-Coder takes approximately 3, 6, and 4 hours. By comparison, all traditional detectors complete repository-level detection within a few minutes per project pair.

\subsubsection{Post-filtering Effectiveness} 

\noindent\textbf{Approach.} We apply SAS filtering to the outputs of the six clone detectors evaluated in RQ1 and assess its effectiveness in identifying code mappings in $D$.

\begin{table}[p]
\centering
\begin{talltblr}[
    theme = mockcaption,
    caption = {Changes in clone detector effectiveness after SAS filtering for CM on dataset $D$.},
    label = {tab:sas-cm},
    remark{Format}={Values are ``$Z$ ($Y$)'', where $Y$ is the filtered result and $Z=Y-X$ is the change from the original result $X$ (Table~\ref{tab:baselines-cm}).},
]{
    width = \columnwidth,
    colspec = {l l *{8}{Q[r,wd=3.5em]}},
    cells = {font=\scriptsize, m},
    colsep = 1.5pt, % 调整列间距，列宽=文字内容的宽度+2*colsep
    rowsep = 1pt, 
    % % --- Merged Cell ---
    cell{1}{3,7} = {c=4}{m},
    cell{1}{1,2} = {r=2}{m}, % pair, tool
    cell{3}{1} = {r=4}{m}, % nicad
    cell{7}{1} = {r=4}{m},
    cell{11}{1} = {r=4}{m},
    cell{15}{1} = {r=4}{m},
    cell{19}{1} = {r=4}{m},
    cell{23}{1} = {r=4}{m},
    % % % --- color ---
    row{even} = {bg=gray!10}, 
    cell{6,10,14,18,22}{2} = {font=\bfseries\scriptsize},
    % column{1,2} = {bg=white},
    row{1,2} = {halign=c, font=\bfseries\scriptsize, bg=gray!30},% header row
    % % --- 线条控制 ---
    hline{1, 3, 27} = {1.2pt, solid}, % 顶线
    hline{2,7,11,15,19,23} = {solid},
    hline{6,10,14,18,22,26} = {solid}, % all pairs
    vline{3,7} = {solid},
}

Tool & Pair & Production Code & &     &    & Test Code &      &      &          \\
     &      &  FPR &  Prec. &  Rec. &  Avg. F1 &  FPR &  Prec. &  Rec. &  Avg. F1  \\

NiCad & \soot/\sootup & \textbf{-0.24} \textcolor{gray}{(0.06)} & \textbf{+0.34} \textcolor{gray}{(0.81)} & -0.02 \textcolor{gray}{(0.56)} & \textbf{+0.139} \textcolor{gray}{(0.768)} & -0.01 \textcolor{gray}{(0.00)} & 0.00 \textcolor{gray}{(0.00)} & 0.00 \textcolor{gray}{(0.00)} & +0.003 \textcolor{gray}{(0.500)} \\
NiCad & \fbugs/\sbugs & -0.07 \textcolor{gray}{(0.06)} & +0.04 \textcolor{gray}{(0.96)} & 0.00 \textcolor{gray}{(0.80)} & +0.027 \textcolor{gray}{(0.850)} & \textbf{-0.14} \textcolor{gray}{(0.03)} & \textbf{+0.34} \textcolor{gray}{(0.85)} & -0.01 \textcolor{gray}{(0.30)} & +0.064 \textcolor{gray}{(0.629)} \\
NiCad & \alrt/\alrf & 0.00 \textcolor{gray}{(0.00)} & +0.08 \textcolor{gray}{(1.00)} & -0.03 \textcolor{gray}{(0.33)} & -0.012 \textcolor{gray}{(0.728)} & -0.12 \textcolor{gray}{(0.04)} & +0.32 \textcolor{gray}{(0.68)} & 0.00 \textcolor{gray}{(0.36)} & \textbf{+0.087} \textcolor{gray}{(0.688)} \\
NiCad & All Pairs & -0.12 \textcolor{gray}{(0.04)} & +0.18 \textcolor{gray}{(0.91)} & -0.01 \textcolor{gray}{(0.69)} & +0.067 \textcolor{gray}{(0.838)} & -0.09 \textcolor{gray}{(0.03)} & \textbf{+0.34} \textcolor{gray}{(0.78)} & -0.01 \textcolor{gray}{(0.32)} & +0.064 \textcolor{gray}{(0.674)} \\
DeepSim & \soot/\sootup & -0.36 \textcolor{gray}{(0.01)} & +0.58 \textcolor{gray}{(0.92)} & -0.06 \textcolor{gray}{(0.35)} & +0.174 \textcolor{gray}{(0.687)} & -0.56 \textcolor{gray}{(0.00)} & 0.00 \textcolor{gray}{(0.00)} & 0.00 \textcolor{gray}{(0.00)} & +0.194 \textcolor{gray}{(0.500)} \\
DeepSim & \fbugs/\sbugs & \textbf{-0.62} \textcolor{gray}{(0.05)} & +0.40 \textcolor{gray}{(0.95)} & 0.00 \textcolor{gray}{(0.50)} & +0.255 \textcolor{gray}{(0.671)} & -0.42 \textcolor{gray}{(0.03)} & \textbf{+0.58} \textcolor{gray}{(0.77)} & -0.02 \textcolor{gray}{(0.17)} & +0.169 \textcolor{gray}{(0.536)} \\
DeepSim & \alrt/\alrf & -0.58 \textcolor{gray}{(0.00)} & \textbf{+0.89} \textcolor{gray}{(1.00)} & -0.10 \textcolor{gray}{(0.43)} & \textbf{+0.406} \textcolor{gray}{(0.783)} & \textbf{-0.75} \textcolor{gray}{(0.06)} & \textbf{+0.58} \textcolor{gray}{(0.79)} & -0.02 \textcolor{gray}{(0.83)} & \textbf{+0.556} \textcolor{gray}{(0.879)} \\
DeepSim & All Pairs & -0.50 \textcolor{gray}{(0.02)} & +0.58 \textcolor{gray}{(0.94)} & -0.02 \textcolor{gray}{(0.45)} & +0.251 \textcolor{gray}{(0.725)} & -0.58 \textcolor{gray}{(0.03)} & +0.62 \textcolor{gray}{(0.78)} & -0.02 \textcolor{gray}{(0.37)} & +0.344 \textcolor{gray}{(0.704)} \\
CCStokener & \soot/\sootup & -0.30 \textcolor{gray}{(0.04)} & +0.43 \textcolor{gray}{(0.87)} & -0.04 \textcolor{gray}{(0.53)} & +0.165 \textcolor{gray}{(0.770)} & -0.04 \textcolor{gray}{(0.00)} & 0.00 \textcolor{gray}{(0.00)} & 0.00 \textcolor{gray}{(0.00)} & +0.009 \textcolor{gray}{(0.500)} \\
CCStokener & \fbugs/\sbugs & -0.27 \textcolor{gray}{(0.14)} & +0.14 \textcolor{gray}{(0.91)} & 0.00 \textcolor{gray}{(0.85)} & +0.121 \textcolor{gray}{(0.852)} & -0.46 \textcolor{gray}{(0.13)} & +0.36 \textcolor{gray}{(0.71)} & -0.02 \textcolor{gray}{(0.53)} & +0.249 \textcolor{gray}{(0.713)} \\
CCStokener & \alrt/\alrf & \textbf{-0.43} \textcolor{gray}{(0.02)} & \textbf{+0.61} \textcolor{gray}{(0.80)} & -0.10 \textcolor{gray}{(0.67)} & \textbf{+0.350} \textcolor{gray}{(0.846)} & \textbf{-0.52} \textcolor{gray}{(0.06)} & \textbf{+0.52} \textcolor{gray}{(0.79)} & -0.02 \textcolor{gray}{(0.87)} & \textbf{+0.393} \textcolor{gray}{(0.892)} \\
CCStokener & All Pairs & -0.34 \textcolor{gray}{(0.06)} & +0.34 \textcolor{gray}{(0.89)} & -0.02 \textcolor{gray}{(0.74)} & +0.188 \textcolor{gray}{(0.851)} & -0.36 \textcolor{gray}{(0.06)} & +0.43 \textcolor{gray}{(0.74)} & -0.02 \textcolor{gray}{(0.64)} & +0.248 \textcolor{gray}{(0.801)} \\
SSCD & \soot/\sootup & -0.07 \textcolor{gray}{(0.08)} & +0.09 \textcolor{gray}{(0.82)} & -0.07 \textcolor{gray}{(0.80)} & +0.018 \textcolor{gray}{(0.864)} & \textbf{-0.48} \textcolor{gray}{(0.00)} & 0.00 \textcolor{gray}{(0.00)} & 0.00 \textcolor{gray}{(0.00)} & +0.156 \textcolor{gray}{(0.500)} \\
SSCD & \fbugs/\sbugs & -0.16 \textcolor{gray}{(0.07)} & +0.08 \textcolor{gray}{(0.96)} & 0.00 \textcolor{gray}{(1.00)} & +0.068 \textcolor{gray}{(0.968)} & -0.36 \textcolor{gray}{(0.13)} & \textbf{+0.27} \textcolor{gray}{(0.81)} & -0.02 \textcolor{gray}{(0.96)} & \textbf{+0.218} \textcolor{gray}{(0.897)} \\
SSCD & \alrt/\alrf & \textbf{-0.26} \textcolor{gray}{(0.02)} & \textbf{+0.51} \textcolor{gray}{(0.81)} & -0.17 \textcolor{gray}{(0.70)} & \textbf{+0.221} \textcolor{gray}{(0.859)} & -0.10 \textcolor{gray}{(0.06)} & +0.21 \textcolor{gray}{(0.79)} & -0.02 \textcolor{gray}{(0.85)} & +0.090 \textcolor{gray}{(0.886)} \\
SSCD & All Pairs & -0.15 \textcolor{gray}{(0.06)} & +0.17 \textcolor{gray}{(0.91)} & -0.03 \textcolor{gray}{(0.91)} & +0.078 \textcolor{gray}{(0.926)} & -0.30 \textcolor{gray}{(0.07)} & +0.37 \textcolor{gray}{(0.80)} & -0.02 \textcolor{gray}{(0.92)} & +0.228 \textcolor{gray}{(0.907)} \\
GPT-OSS & \soot/\sootup & \textbf{-0.08} \textcolor{gray}{(0.03)} & +0.14 \textcolor{gray}{(0.92)} & -0.09 \textcolor{gray}{(0.69)} & \textbf{+0.015} \textcolor{gray}{(0.854)} & 0.00 \textcolor{gray}{(0.00)} & 0.00 \textcolor{gray}{(0.00)} & 0.00 \textcolor{gray}{(0.00)} & 0.000 \textcolor{gray}{(0.500)} \\
GPT-OSS & \fbugs/\sbugs & -0.02 \textcolor{gray}{(0.00)} & +0.01 \textcolor{gray}{(1.00)} & 0.00 \textcolor{gray}{(0.93)} & +0.008 \textcolor{gray}{(0.956)} & 0.00 \textcolor{gray}{(0.00)} & +0.01 \textcolor{gray}{(0.99)} & -0.02 \textcolor{gray}{(0.81)} & -0.004 \textcolor{gray}{(0.916)} \\
GPT-OSS & \alrt/\alrf & -0.02 \textcolor{gray}{(0.00)} & \textbf{+0.24} \textcolor{gray}{(1.00)} & -0.17 \textcolor{gray}{(0.27)} & -0.067 \textcolor{gray}{(0.686)} & \textbf{-0.02} \textcolor{gray}{(0.01)} & \textbf{+0.07} \textcolor{gray}{(0.93)} & -0.08 \textcolor{gray}{(0.70)} & -0.008 \textcolor{gray}{(0.877)} \\
GPT-OSS & All Pairs & -0.04 \textcolor{gray}{(0.01)} & +0.07 \textcolor{gray}{(0.98)} & -0.04 \textcolor{gray}{(0.81)} & +0.009 \textcolor{gray}{(0.910)} & -0.01 \textcolor{gray}{(0.01)} & +0.03 \textcolor{gray}{(0.97)} & -0.04 \textcolor{gray}{(0.77)} & -0.004 \textcolor{gray}{(0.912)} \\
Qwen-Coder & \soot/\sootup & 0.00 \textcolor{gray}{(0.00)} & 0.00 \textcolor{gray}{(0.98)} & -0.02 \textcolor{gray}{(0.43)} & -0.013 \textcolor{gray}{(0.741)} & 0.00 \textcolor{gray}{(0.00)} & 0.00 \textcolor{gray}{(0.00)} & 0.00 \textcolor{gray}{(0.00)} & 0.000 \textcolor{gray}{(0.500)} \\
Qwen-Coder & \fbugs/\sbugs & 0.00 \textcolor{gray}{(0.00)} & 0.00 \textcolor{gray}{(1.00)} & 0.00 \textcolor{gray}{(0.88)} & 0.000 \textcolor{gray}{(0.926)} & 0.00 \textcolor{gray}{(0.00)} & 0.00 \textcolor{gray}{(1.00)} & -0.02 \textcolor{gray}{(0.46)} & -0.010 \textcolor{gray}{(0.747)} \\
Qwen-Coder & \alrt/\alrf & 0.00 \textcolor{gray}{(0.00)} & 0.00 \textcolor{gray}{(1.00)} & 0.00 \textcolor{gray}{(0.27)} & 0.000 \textcolor{gray}{(0.686)} & 0.00 \textcolor{gray}{(0.01)} & 0.00 \textcolor{gray}{(0.92)} & 0.00 \textcolor{gray}{(0.62)} & 0.000 \textcolor{gray}{(0.845)} \\
Qwen-Coder & All Pairs & 0.00 \textcolor{gray}{(0.00)} & 0.00 \textcolor{gray}{(1.00)} & -0.01 \textcolor{gray}{(0.70)} & -0.004 \textcolor{gray}{(0.865)} & 0.00 \textcolor{gray}{(0.01)} & 0.00 \textcolor{gray}{(0.97)} & -0.01 \textcolor{gray}{(0.51)} & -0.006 \textcolor{gray}{(0.799)} \\
\end{talltblr}
\end{table} 

\noindent\textbf{Results.} Table~\ref{tab:sas-cm} reports the changes in detection results and the post-filtering values. For all four traditional detectors, filtering consistently reduces the false positive rate and increases precision, indicating that SAS effectively removes irrelevant clone pairs. The improvement is larger for detectors with lower initial precision: DeepSim, the weakest traditional detector in RQ1, benefits the most, with precision reaching 0.94 (production) and 0.78 (test) at the ``All Pairs'' level, whereas NiCad, with the lowest initial FPR, shows more modest gains.

These precision gains come with a modest and expected recall cost. Nevertheless, the net effect, measured by average F1, remains positive for all four traditional detectors at the ``All Pairs'' level (NiCad +0.067/+0.064, DeepSim +0.251/+0.344, CCStokener +0.188/+0.248, SSCD +0.078/+0.228, production/test respectively), indicating that the precision gains outweigh the recall losses. For NiCad, the average F1 gain is more modest than the precision improvement because average F1 also includes the non-CM class, whose F1 is already high before filtering. This reflects a property of the aggregate metric rather than a weaker filtering effect.

After SAS filtering, the strongest traditional detector, SSCD, achieves average F1 of 0.926/0.907 on production/test code, comparable to or higher than the LLM-based baselines: GPT-OSS achieves 0.910/0.912 and Qwen-Coder achieves 0.865/0.799. CCStokener also achieves comparable results to Qwen-Coder, particularly on test code (0.801 vs. 0.799). These results demonstrate that \emph{SAS enables strong traditional clone detectors to achieve CM detection effectiveness comparable to LLM detectors, while avoiding LLM inference costs}. In contrast, SAS provides little additional benefit for the LLM detectors, whose initial FPRs are already low, resulting in negligible or slightly negative changes in average F1.

To assess the statistical significance of these changes, we conducted McNemar's tests~\citep{McNemar1947} on the samples covered by each detector in $D$ (575 code mappings and 1,230 non-mappings), comparing predictions before and after SAS filtering. The reduction in false positives was significant for all detectors ($p<0.001$) except Qwen-Coder, which produced no false positives. The recall decrease was significant for GPT-OSS, DeepSim, CCStokener, and SSCD ($p<0.001$), but not for NiCad or Qwen-Coder ($p=0.0625$).

\noindent\textbf{Post-Filtering Baseline Detector Reports.} To assess the practical reduction in candidate pairs at the scale of actual detector reports, we applied the universal SAS threshold of $t=0.70$ to the results reported by the five baseline detectors. Since this analysis estimates filtering effectiveness at realistic report scale rather than repeating the fine-grained evaluation, we report results aggregated across all code types. Table~\ref{tab:postfilt-raw} reports the total and retained pairs and the percentage removed after filtering. DeepSim reports substantially more pairs than the other detectors, consistent with its pairwise embedding-similarity computation over a broad comparison space. SAS removes over 99.2\% of DeepSim's reported pairs across all three project pairs, while also substantially reducing the reports of the other detectors, thereby reducing the number of pairs requiring further inspection.

% Requires: \usepackage{tabularray}
\begin{table}[!t]
\centering
\caption{Effectiveness of SAS filtering ($t=0.70$) on full tool reports for CM.}
\label{tab:postfilt-raw}
\begin{tblr}{
  width = \textwidth,
  colspec = {l r r r r r r r r r},
  cells = {m, font=\footnotesize},
  colsep = 2.5pt, % 调整列间距，列宽=文字内容的宽度+2*colsep
  rowsep = 2pt, 
  row{1,2} = {halign=c, font=\bfseries\footnotesize, bg=gray!30},% header row
  cell{1}{1} = {r=2}{m},
  cell{1}{2} = {c=3}{m},
  cell{1}{5} = {c=3}{m},
  cell{1}{8} = {c=3}{m},
  % % --- 线条控制 ---
  hline{1,3,9} = {1.2pt, solid}, % 顶线
  hline{2} = {solid},
  vline{2,5,8} = {solid},
}

Tool    & \soot/\sootup &      &            & \fbugs/\sbugs &      &           & \alrt/\alrf &   & \\
        & Total         & Kept & Drop (\%)  & Total         & Kept & Drop (\%) & Total       & Kept & Drop (\%)   \\
        
NiCad & 514 & 147 & 71.40 & 2,119 & 1,958 & 7.60 & 65 & 38 & 41.54 \\
DeepSim & 2,092,140 & 734 & 99.96 & 1,203,587 & 9,606 & 99.20 & 317,406 & 141 & 99.96 \\
CCStokener & 2,118 & 241 & 88.62 & 4,234 & 2,810 & 33.63 & 10,746 & 157 & 98.54 \\
SSCD & 11,017 & 491 & 95.54 & 21,133 & 7,256 & 65.67 & 4,650 & 150 & 96.77 \\
GPT-OSS & 221 & 136 & 38.46 & 3,829 & 3,634 & 5.09 & 65 & 48 & 26.15 \\
Qwen-Coder & 86 & 75 & 12.79 & 3,546 & 3,460 & 2.43 & 52 & 51 & 1.92 \\

\end{tblr}
\end{table}

\noindent\textbf{SAS Computation Time Overhead.} We measured SAS computation time on the union of method pairs reported by all five baseline detectors on an Apple M3 machine (8-core CPU, 20GB RAM). The union contains 2.10M, 1.21M, and 324K pairs for \sootup/\soot, \fbugs/\sbugs, and \alrt/\alrf, respectively, with DeepSim alone accounting for over 97\% of the pairs in each union. The corresponding SAS computation times are 1,638s, 772s, and 172s, yielding a throughput of approximately 1,280--1,890 pairs/s. Using this throughput and the report sizes in Table~\ref{tab:postfilt-raw}, SAS takes at most 5.7s for the four non-DeepSim baselines, whereas DeepSim requires an estimated 168--1,634s. Given its FPR of up to 0.75 in Table~\ref{tab:baselines-cm}, DeepSim's substantially larger reports account for its higher SAS overhead; for the other baselines, the overhead remains small.

\subsubsection{Robustness of SAS Weight Tuning}
\label{ssubsec:stability}
         % Approach.（5-fold StratifiedKFold 协议、双目标 grid search、
         % per-fold threshold sweep）
         % Results.（mean±std/range 表、"上游波动不传导下游"论证、
         % 全量参数 vs CV 均值一致性检验）
         
\noindent \textbf{Approach.} 
To assess the robustness of SAS weight tuning to variations in the labeled samples, we perform 5-fold cross-validation on the labeled production-code pairs of \soot/\sootup (Table~\ref{tab:samples}). Folds are stratified by the three mutually exclusive pair labels (non-clone, irrelevant clone, and code mapping) to preserve their distributions across splits. Component similarities are precomputed once for all pairs and reused across folds, ensuring that variation between folds are caused only by the data partition used for tuning. In each iteration, four folds are used to select weights via the same procedure as Section~\ref{subsec:post-filtering}, with R-precision over code mappings as the optimization objective. The selected weights are then applied to the remaining fold to characterize their filtering performance. Specifically, the decision threshold is swept over [0, 1] with a step size of 0.05, selecting the threshold that maximizes average F1 on the held-out fold. In case of ties, the larger threshold is selected because it yields a more conservative decision boundary. For each fold, we record the selected weights, the R-precision used for weight selection, the resulting optimal threshold, and the corresponding maximum average F1 on the held-out fold.  Finally, we repeat the procedure on the full training samples to obtain the final weight configuration reported earlier.

\begin{table}[!t]
\centering
\begin{talltblr}[
    theme = mockcaption,
    caption = {Cross-validation results for the CM objective on the production-code samples of \soot/\sootup in $D$.},
    label = {tab:cv_robustness},
    note{1}={\textit{Abs. Diff.} denotes $|\text{All Samples} - \text{Mean}|$ for each column.},
    note{2}={Mean weights evaluated on full samples: R-Precision = 0.85.},
]{
    width = \columnwidth,
    colspec = {lrrrrrrrrr},
    % --- Merged Cell ---
    cell{1}{2} = {c=7}{c}, % Train 跨 7 列居中
    cell{1}{1} = {r=2}{m},
    cell{1}{9} = {c=2}{c}, % Test 跨 2 列居中
    % cell{10}{1} = {c=10}{c}, % 比较章节标题跨 10 列靠左
    % % --- color ---
    row{1,2} = {font=\bfseries, bg=gray!30},
    % --- 线条控制 ---
    hline{1, 10, 12} = {1.2pt, solid}, % 顶线
    hline{2,3,8,11} = {solid},
    vline{9} = {1pt, solid}, % train 和 test 分割线
    vline{2,5,8,9} = {solid},
}
Fold & Train Fold & & & & & & & Test Fold & \\
 & $\alpha$ & $\beta$ & $\theta$ & $\delta$ & $\eta$ & $\phi$ & R-Prec. & Thres. & Avg. F1 \\

1 & 0.15 & 0.75 & 0.10 & 0.90 & 0.05 & 0.05 & 0.87 & 0.80 & 0.88 \\
2 & 0.10 & 0.70 & 0.20 & 0.80 & 0.10 & 0.10 & 0.86 & 0.70 & 0.90 \\
3 & 0.15 & 0.80 & 0.05 & 0.75 & 0.05 & 0.20 & 0.86 & 0.75 & 0.88 \\
4 & 0.15 & 0.80 & 0.05 & 0.75 & 0.10 & 0.15 & 0.86 & 0.80 & 0.89 \\
5 & 0.05 & 0.70 & 0.25 & 0.70 & 0.05 & 0.25 & 0.88 & 0.60 & 0.85 \\
SD & 0.04 & 0.05 & 0.09 & 0.08 & 0.03 & 0.08 & 0.01 & 0.08 & 0.02 \\
Mean \TblrNote{2}   & 0.12 & 0.75 & 0.13 & 0.78 & 0.07 & 0.15 & 0.87 & 0.73 & 0.88 \\
All Samples         & 0.05 & 0.70 & 0.25 & 0.70 & 0.05 & 0.25 & 0.87 & --   & -- \\
Abs. Diff. \TblrNote{1} 
                    & 0.07 & 0.05 & 0.12 & 0.08 & 0.02 & 0.10 & 0.00 & -- & -- \\
\end{talltblr}
\end{table}

\noindent \textbf{Results.}
Table~\ref{tab:cv_robustness} summarizes the results of the 5-fold cross-validation. For each fold, the table reports the learned weight configurations and corresponding tuning outcomes, together with their mean and standard deviation (SD) across folds.  It also includes the configuration obtained by applying the same procedure to all samples and the absolute deviation (Abs. Diff.) between the all-sample values and the cross-validated means.

Across the five folds, the learned weights exhibit some variation, with standard deviations ranging from 0.03 to 0.09. This variation partly reflects the compositional constraints noted in Section~\ref{subsec:post-filtering}, whereby an increase in one weight within a group is necessarily offset by the others. Thus, the variation represents redistributions among correlated weights. \emph{Despite this variation in individual weights, the downstream outcomes remain relatively robust}, with standard deviations of 0.01 for R-Precision, 0.08 for the selected thresholds, and 0.02 for held-out average F1 across folds.

The weights used in our main evaluation, which were tuned on all samples, differ from the cross-validated means by up to 0.12 across the six weights. To assess whether these differences affect filtering quality, we additionally evaluated the cross-validated mean weights on all samples, obtaining an R-Precision of 0.85, compared with 0.87 for the all-sample configuration. This small difference suggests that \emph{the filtering outcome is relatively insensitive to the exact weight configuration}. We therefore use the all-sample configuration for the main evaluation, as it results from the standard tuning procedure, rather than post hoc selection.

\subsubsection{Threshold Generalizability Across Projects}
\label{ssubsec:threshold}
 
\noindent \textbf{Approach.} We further examine whether SAS thresholds generalize across project pairs, or whether the method requires project-specific calibration to remain effective. We conduct a SAS threshold sweep for each project pair over all production and test code, which corresponds to the deployment setting of our approach. From the sweep results, we identify the \emph{oracle threshold} for each pair as the threshold achieving the highest average F1. We then derive a \emph{universal threshold} by intersecting the tolerance regions of the three pairs. Each tolerance region contains thresholds whose average F1 is within the specified tolerance of the pair's oracle F1. We consider tolerance levels from 0.99 to 0.80 and use the first level at which the three regions overlap. When the intersection contains multiple thresholds, we select its lower bound as the universal threshold. This procedure is applied separately to the two detection objectives, genuine clone (GC) and code mapping (CM). Finally, we compare the average F1 at the universal threshold with the corresponding oracle F1 for each project pair.

\begin{figure}[!t]
    \centering
    \begin{subfigure}[t]{0.48\textwidth}
        \centering
        \includegraphics[width=\textwidth]{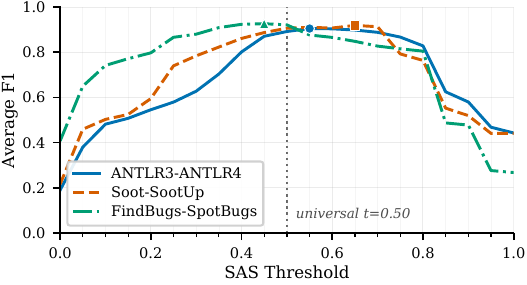}
        \caption{GC objective}
        \label{fig:thres-gc}
    \end{subfigure}
    \hfill
    \begin{subfigure}[t]{0.48\textwidth}
        \centering
        \includegraphics[width=\textwidth]{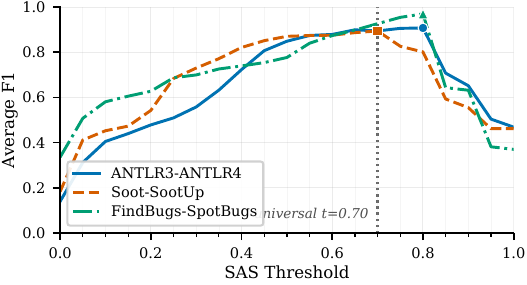}
        \caption{CM objective}
        \label{fig:thres-cm}
    \end{subfigure}

    \caption{Average F1 across SAS thresholds for each project pair.}
    \label{fig:thres-gen}
\end{figure}

\noindent \textbf{Results.}
Figure~\ref{fig:thres-gen} shows the average F1 across SAS thresholds for the three project pairs. Across both objectives, the curves remain relatively flat over a broad range of thresholds, and the selected universal threshold falls within the high-F1 region of all three pairs. This indicates that moderate deviations from a pair-specific oracle threshold generally have limited impact on F1, allowing a common threshold to be selected across project pairs.
Table~\ref{tab:thres-gen} compares the oracle and universal thresholds. For GC, the universal threshold is $t=0.50$, with F1 losses of $0.005$--$0.013$ across pairs. For CM, the universal threshold is $t=0.70$, resulting in F1 losses of $0$--$0.043$ across the three pairs. The largest loss occurs for \fbugs/\sbugs, but its average F1 remains high at $0.926$. Thus, \emph{the universal thresholds retain high average F1 across project pairs}.

\begin{table}[!t]
    \centering
    \caption{Comparison of average F1 at the universal and oracle thresholds across project pairs on dataset $D$.}
    \label{tab:thres-gen}

    \begin{tblr}{
        width = \columnwidth,
        colspec = {l l r r r r r},
        cells = {m},
        colsep = 4pt, % 调整列间距，列宽=文字内容的宽度+2*colsep
        % rowsep = 2.5pt, % default: 2pt
        % --- Merged Cell ---
        cell{2,5}{1} = {r=3}{m},
        % --- 线条控制 ---
        hline{1,2,Z} = {1.2pt, solid},
        hline{5} = {solid},
        vline{2,3,5,7} = {solid},
        % --- color ---
        row{1} = {font=\bfseries, bg=gray!30},
    }
    
Objective & Pair & Oracle t & Oracle F1 & Universal t & Universal F1 & $\Delta$F1 \\
CM & \soot/\sootup & 0.70 & 0.894 & 0.70 & 0.894 & 0.000 \\
CM & \fbugs/\sbugs & 0.80 & 0.970 & 0.70 & 0.926 & -0.043 \\
CM & \alrt/\alrf & 0.80 & 0.907 & 0.70 & 0.894 & -0.013 \\
GC & \soot/\sootup & 0.65 & 0.919 & 0.50 & 0.906 & -0.013 \\
GC & \fbugs/\sbugs & 0.45 & 0.926 & 0.50 & 0.921 & -0.005 \\
GC & \alrt/\alrf & 0.55 & 0.905 & 0.50 & 0.892 & -0.013 \\
    \end{tblr}
\end{table}

\begin{tcolorbox}[left=0pt,right=0pt,top=0pt,bottom=0pt]
\textbf{Finding 7:} 
SAS filtering substantially improves code mapping detection for traditional clone detectors by reducing false positives, yielding gains of up to 0.62 in precision and 0.34 in average F1 with all project pairs pooled. With SAS, strong traditional detectors can achieve CM detection effectiveness comparable to LLM detectors. Tuned weights yield robust downstream filtering outcomes across labeled samples, while a universal threshold generalizes across project pairs.
\end{tcolorbox}

\subsection{RQ3: Efficacy of Individual Rules}
\label{subsec:rq5}

\noindent\textbf{Approach.}
To assess the contribution of individual rules in Table~\ref{tab:pattern-rule}, we exclude each rule by removing its corresponding code information from SAS computation (Equation~\ref{eq:component}), resulting in five settings: \textbf{ALL} (all rules applied), \textbf{EXR1} (R1 excluded by disabling custom renaming rules in preprocessing), \textbf{EXR2} (R2 excluded with $simLocalVar = simMethodHeader = 0$), \textbf{EXR3} (R3 excluded with $simMethodDoc = simClassDoc = 0$), and \textbf{EXR4} (R4 excluded with $simComment = 0$).
We conduct all analyses on the entire dataset $D$ without distinguishing project pairs or code types. For each setting, we evaluate detection performance using the maximum and mean average F1 across thresholds, and ranking changes using the number and percentage of affected samples and the maximum and median changes in SAS and rank relative to \textbf{ALL}.

\begin{table}[!t]
    \centering
    \begin{talltblr}[
        theme = mockcaption,
        caption = {Average F1 performance and impact of individual SAS rules on $D$}, 
        label = {tab:idv-rule},
        remark{Notation}={Max/Mean F1 denote the maximum/mean average F1 across thresholds; Thres. is the threshold for Max F1; Affected and Cov. denote the number and percentage of affected samples; Max/Med. $\Delta$SAS and Max/Med. $\Delta$Rank denote the maximum/median changes in SAS and rank, respectively.},
    ]{
        width = \columnwidth,
        colspec = {l *{9}{r}},
        cells = {font=\scriptsize},
        colsep = 2.5pt, % 调整列间距，列宽=文字内容的宽度+2*colsep
        rowsep = 2pt, 
        % % --- merge cell ---
        cell{6}{7} = {c=4}{halign=c},
        % % --- color ---
        row{1} = {font=\bfseries\scriptsize, bg=gray!30},% header row
        % % % --- 线条控制 ---
        hline{1, 2, 7} = {1.2pt, solid}, % 顶线
        hline{6} = {solid},
        vline{2,5,7,9} = {solid},
    }

    Setting & Max F1 & Thres. & Mean F1 
            & Affected & Cov.(\%) & Max $\Delta$SAS & Med. $\Delta$SAS & Max $\Delta$Rank & Med. $\Delta$Rank  \\
            
    EXR1 & 0.923 & 0.75 & 0.691 & 80 & 4.43 & 0.13 & 0.03 & 312 & 52 \\
    EXR2 & 0.808 & 0.05 & 0.432 & 1662 & 92.08 & 0.78 & 0.42 & 1212 & 241 \\
    EXR3 & 0.924 & 0.80 & 0.681 & 634 & 35.12 & 0.10 & 0.01 & 380 & 20 \\
    EXR4 & 0.929 & 0.75 & 0.679 & 368 & 20.39 & 0.08 & 0.08 & 345 & 67 \\
    ALL & 0.926 & 0.80 & 0.691  & 1734 & 96.07 & --- & & &  \\

    % EXR1 & 0.952 & 0.85 & 0.729 & 78    & 3.87  & 0.15 & 0.04 & 463 & 34 \\
    % EXR2 & 0.833 & 0.15 & 0.462 & 1832  & 90.78 & 0.78 & 0.35 & 1211 & 211 \\
    % EXR3 & 0.958 & 0.85 & 0.726 & 682   & 33.80 & 0.10 & 0.02 & 448 & 46 \\
    % EXR4 & 0.957 & 0.85 & 0.723 & 376   & 18.63 & 0.03 & 0.03 & 329 & 28 \\
    % ALL  & 0.956 & 0.85 & 0.730 & 1924  & 95.34 & --- & & &  \\

    \end{talltblr}
\end{table}

\noindent\textbf{Results.}
Table~\ref{tab:idv-rule} shows that \emph{R2 (similar identifier name matching) has the broadest coverage and the largest SAS impact}. EXR2 affects 92.08\% of samples in $D$, with a median SAS change of 0.42 among affected samples. Excluding R2 reduces the maximum average F1 from 0.926 to 0.808 and the mean average F1 from 0.691 to 0.432. This effect is partly expected because its corresponding \textit{simMethodHeader} component receives the largest weight ($\beta=0.75$).

\emph{The other rules have narrower but heterogeneous coverage and still alter the ranking of affected samples.} R1, R3, and R4 affect 4.43\%, 35.12\%, and 20.39\% of samples, respectively, with median rank changes of 52, 20, and 67 positions. Thus, although their coverage is substantially lower than R2's, excluding each rule changes the relative ordering of their affected samples. Their effects on average F1 are small: maximum F1 varies by at most 0.003, while mean F1 remains within 0.679--0.691 across exclusion settings, compared with 0.691 under ALL.

\begin{tcolorbox}[left=0pt,right=0pt,top=0pt,bottom=0pt]
\textbf{Finding 8:}
The ``R2: Similar Identifier Name Matching'' rule affects the most samples in our dataset and shifts their rankings the most, while the other rules affect smaller subsets but still alter their relative rankings.
\end{tcolorbox}

\subsection{RQ4: Comparison of SAS-Filtered and Task-Aligned LLM Detection}

\begin{table}[!t]
\centering
\begin{talltblr}[
        theme = mockcaption,
        caption = {Comparison of GPT-OSS with a CM prompt and SAS-filtered traditional detectors on dataset $D$ (All Pairs).}, 
        label = {tab:llm4cm},
    ]{
    width = \columnwidth,
    colspec = {X[2,l] *{3}{X[r]} X[2,r] *{3}{X[r]} X[2,r]},
    cells = {font=\footnotesize, m},
    % colsep = 3pt, % 调整列间距，列宽=文字内容的宽度+2*colsep
    rowsep = 2pt, 
    % % % --- Merged Cell ---
    cell{1}{2,6} = {c=4}{m},
    cell{1}{1} = {r=2}{m}, %  tool
    row{1,2} = {halign=c, font=\bfseries\footnotesize, bg=gray!30},% header row
    % row{18,19,20,21,22} = {bg=gray!10},
    % % % --- 线条控制 ---
    hline{1, 3, 6} = {1.2pt, solid}, % 顶线
    hline{2} = {solid},
    vline{2,6} = {solid},
    %% --- bold font ---
    cell{3}{2,4,6,8} = {font=\bfseries\footnotesize},
    cell{5}{3,5,7,9} = {font=\bfseries\footnotesize},
}

Tool & Production Code & &     &    & Test Code &      &      &          \\
     & FPR & Prec. & Rec. & Avg. F1 & FPR & Prec. & Rec. & Avg. F1  \\

GPT-OSS     & 0.13 & 0.82 & 0.98 & 0.909 & 0.12 & 0.71 & 0.98 & 0.878 \\
CCStokener  & 0.06 & 0.89 & 0.74 & 0.851 & 0.06 & 0.74 & 0.64 & 0.801 \\
SSCD        & 0.06 & 0.91 & 0.91 & 0.926 & 0.07 & 0.80 & 0.92 & 0.907 \\
    \end{talltblr}
\end{table}

\noindent\textbf{Approach.} In RQ1 and RQ2, we evaluate clone detectors for code mapping, with LLM detectors using the GC prompt. To provide a task-aligned comparison, we evaluate GPT-OSS with the CM prompt without SAS filtering. We compare it with SAS-enhanced traditional detectors that performed close to LLM detectors in RQ2, including CCStokener and SSCD. The results for CCStokener and SSCD are reproduced from Table~\ref{tab:sas-cm}, while the GPT-OSS results are newly obtained for this RQ. Both configurations are reported at the All Pairs level, separately for production and test code.

\noindent\textbf{Results.} As shown in Table~\ref{tab:llm4cm}, both SAS-enhanced traditional detectors achieve lower FPR and higher precision than GPT-OSS with the CM prompt. SSCD with SAS filtering achieves a higher average F1 than GPT-OSS on both production and test code (0.926 vs. 0.909 and 0.907 vs. 0.878, respectively). CCStokener also benefits from reduced false positives, but its lower recall limits its F1 relative to GPT-OSS.

\begin{tcolorbox}[left=0pt,right=0pt,top=0pt,bottom=0pt]
\textbf{Finding 9:} 
SAS-enhanced traditional clone detectors can approach or surpass LLM-based detectors with a CM-oriented prompt, with SSCD outperforming GPT-OSS.
\end{tcolorbox}

\section{Threats to Validity}

\noindent \textbf{Internal.}
Manual labeling may introduce subjectivity when distinguishing code mappings, genuine clones, irrelevant clones, and non-clones. To reduce this threat, two authors independently labeled the sampled pairs and resolved disagreements through discussion, achieving substantial to almost perfect inter-rater agreement for most subsets (Table~\ref{tab:kappa}). The SAS weights may overfit the \soot/\sootup production-code samples used for tuning. We assessed this risk through five-fold cross-validation on these samples, which showed robust held-out F1 despite variation in the tuned weights (Table~\ref{tab:cv_robustness}).  Similarly, the selected threshold may be specific to an individual project pair. We therefore compared pair-specific oracle thresholds with a universal threshold, with at most 0.043 F1 loss (Table~\ref{tab:thres-gen}), while retaining high average F1 across project pairs. Finally, SAS's effectiveness may be reduced by non-systematic renaming that disrupts textual continuity. We mitigate this risk by allowing project-specific renaming rules to be manually specified (R1). The completeness of our keyword- and label-based search for redesign-related artifacts (Section~\ref{sec:study}) was not systematically validated, and links using other terminology may have been missed.

\noindent \textbf{External.}
Our detector-based sampling yields uneven sample sizes and compositions, as some detectors report fewer candidates or candidates dominated by non-clones and false positives. This sampling reflects each detector's actual output rather than a balanced instance pool, so subset-level estimates should be interpreted with awareness of this variation. Because the benchmark is constructed from detector outputs, it does not cover code mappings missed by all evaluated detectors and therefore does not assess the ability to discover such mappings. Type-4 clones are underrepresented in our labeled dataset (Table~\ref{tab:samples}), as some detectors are less effective at detecting them. Moreover, our results may not generalize beyond the three evaluated redesigned pairs, selected clone detection techniques, and Java-based projects. To enhance general applicability, we derived the rules from the \soot/\sootup redesign but evaluated them on two held-out redesigned pairs, \fbugs/\sbugs and \alrt/\alrf, and considered three diverse traditional clone detection techniques and an LLM-based technique instantiated with two different LLMs.

% \noindent \textbf{External.}
% Our approach improves efficacy across four evaluated tools.
% As the efficacy may vary with other tools, we selected diverse clone detection  techniques to enhance representation.
% Our approach is effective on the evaluated samples but may vary on other datasets.
% To mitigate this, we evaluated it on a fairly large dataset of over 385 samples to ensure a 95\% confidence level.
% Moreover, our results may not generalize beyond \soot/\sootup and \fbugs/\sbugs.
% % To ensure the general applicability of our findings and validate our manual analysis,
% To enhance general applicability,
% we (1) derived our rules from the bidirectional study on \soot/\sootup and evaluated on two project pairs, and (2) reported any faulty clone pairs to the developers.

% \noindent \textbf{Internal.}
% We manually compared each clone pair for correctness. To reduce potential bias, two authors independently verified the labels and resolved any conflicts through discussion.

\section{Related Work}
\noindent\textbf{Studies.} Prior work has examined multiple versions of a software project~\citep{linuxbackport,nversion}, e.g., developers' experiences in redesigning their own projects~\citep{dorman2001redesign,karakayasootup}.
Our study can be seen as a special case of N-version programming, e.g., a study examined N-version programming where two research teams add floating-point support for KLEE independently~\citep{nversion}.
Another study focuses on patch backporting in the Linux kernel~\citep{linuxbackport}. %where commits submitted to Linux mainline are backported to older versions~\citep{linuxbackport}. 
Our study differs from prior studies in two key aspects: (1) we study redesigned projects instead of program versions,
and (2) we focus on code and test reuse practices instead of other practices (e.g., adding new support~\citep{nversion} or backporting patches~\citep{linuxbackport}).

\noindent\textbf{Code clone detection.} 
Although some code clone detection tools target specific domains (e.g., smart contracts~\citep{he2020characterizing}, apps~\citep{wang2015wukong}, pre-training~\citep{10.1145/3597926.3598035}, or clone synchronization~\citep{7503722}), we focus on general-purpose code clone detection tools introduced in Section~\ref{introclone}.
In the future, it is worthwhile to study how to enhance domain-specific clone detection for redesigned projects.
Prior approaches typically rely on abstract representations (e.g., program dependence graphs) or code normalization to capture similarity, but often ignore identifier names and code comments, which are crucial in redesigned projects. Documentation contains valuable domain knowledge, yet many tools ignore it due to the ambiguity of natural language.  Several studies have confirmed its usefulness for code clone detection~\citep{kuttal2020source,gupta2021identifying}, e.g., Kuttal et al~\citep{kuttal2020source} use Latent Dirichlet Allocation (LDA) at file level combined with PDGs for function-level detection, while Gupta and Goyal~\citep{gupta2021identifying} extract method documentation and apply Latent Semantic Indexing (LSI) with various similarity measures to identify concept-level clones. In contrast, our approach leverages near-duplicate documentation in redesigned projects using a simpler LCS-based method~\citep{koznov2024calculating}, and combines it with identifier information. Moreover, our rules can be applied to prior code clone detection tools for further effectiveness enhancement. 

\noindent\textbf{Identifier Similarity.} Prior work shows that combining identifier similarity with structural information is effective for clone detection. Marcus et al. \citep{idf2001} and CLAN \citep{apicall} use LSI to capture conceptual similarity, using code comments and identifiers or API calls from common libraries. Higo et al. \citep{vocabulary} show that combining vocabulary similarity with structural similarity reduces false positives, but vocabulary-based measures are less effective for method pairs within the same file, where methods often share fields. Our SAS is related to vocabulary similarity but is tailored to redesigned projects. Specifically, we (1) model identifiers in a finer-grained manner, e.g., decomposing signature similarity into return type, method name, and parameters; (2) preserve all identifier tokens instead of filtering by part of speech or stop words, since identifiers are short and often reused or slightly modified during redesign; and (3) use LCS rather than Jaccard similarity to retain token order, which is usually preserved under local modifications and helps mitigate limitations of vocabulary similarity for intra-file methods.

\noindent\textbf{Code reuse.} Researchers have explored code reuse to avoid re-implementing functionality~\citep{Gharehyazie2017,wang2016hunter,haefliger2008code}, including techniques for reusing low-level code~\citep{reuse2,reuse1}. Recent work has also studied the characteristics of reusable code clones using large-scale empirical analysis~\citep{studyofclones}. Unlike prior work, we (1) propose rules to improve clone detection and establish code mappings between redesign and original projects, and (2) focus on redesigned projects rather than general  cross-project clones.

\section{Lessons Learned}

We discuss the lessons learned and implications based on our study.

\noindent\textbf{Ongoing redesign projects should consider bidirectional porting instead of focusing on forward porting of code.} While our study began by analyzing reuse from \soot$\rightarrow$\sootup, we observed that reuse is needed in the reverse direction (\sootup$\rightarrow$\soot). 
\sootup developers tend to reuse production code more often than test code (Finding 2), whereas \soot can benefit from reusing \sootup’s tests (C2). 
Beyond reporting findings, we contributed to the reuse effort by porting fixes and tests to both \soot and \sootup.
For \fbugs/\sbugs, bidirectional porting is unnecessary as \fbugs is deprecated.
Unlike \fbugs's developers, \soot is still being actively maintained, suggesting that bidirectional porting between \soot and \sootup is needed. While \emph{developers of ongoing redesign projects may get bogged down in code migration, adopting the Linux kernel’s well-defined backporting approach~\citep{linuxbackport,coccinelle,backport} can help}.
As both projects evolve, our proposed rules and techniques help reduce the manual effort to identify code mappings.

\noindent\textbf{Beyond static analyzers.}
Although our study focuses on static analyzers (a challenging cross-repo scenario where redesign needs often arise), the implications of several findings may extend to other cross-repository redesign settings. Our action research also translates these findings into practice through open-source contributions. Moreover, most of our proposed rules are general (except for R1 requiring redesign-specific renaming rules) and were validated on held-out redesigned projects. Thus, they can be applied to enhance existing traditional clone detection techniques to uncover the code mappings between the redesigned projects. 
% Moreover, as most of our proposed rules are general (except for R1 where users need to add customized replacement rules), they can be directly applied to enhance existing clone detection techniques to uncover the code mappings between the redesigned projects.  

\noindent\textbf{Automated change tracking in redesigned projects.}
As the migration to \sootup is still an ongoing effort, developers must maintain both versions until \sootup becomes ``feature-complete''. When both projects evolve, their evolution leads to divergence in code and tests, scattering related changes across multiple artifacts (Finding 1).
To find all relevant changes, we need to conduct both keyword and label searches across source code, commits, issues/PRs, and documentation,
indicating the need for automated tools to track relevant changes during redesign.
One possible solution is adding a GitHub feature~\citep{crossprojectpr} to support linking a PR or issue to multiple projects, but it only partially solves the problem as it misses relevant discussions in documentation. 
Hence, tools that can track relevant changes, focusing on the mapping of changes between the original and redesigned projects, are a worthwhile direction to explore. Our rules, which can be applied on top of prior code clone detection tools, can help re-establish these links.

\noindent\textbf{Technical debts as deferred code reuse.}
Our study revealed that technical debt exists in TODO comments (C1) as a hidden form of deferred code reuse. Adapting the code within the TODOs often required substantial effort because the TODOs contained partially completed and outdated code.
In future, it is worthwhile to design code generation techniques that can automatically reuse and adapt code.

\noindent\textbf{Test reuse for static analyzers.} Our study reveals missed opportunities for backporting tests between the two static analysis projects, as developers often rewrite tests instead of reusing them, increasing the maintenance burden.  While several automated testing approaches for static analyzers~\citep{austin2023ecstatic,zhang2023statfier,he2024finding,zhang2024annatester,zhang2024sascope,kaindlstorfer2024interrogation,10.1145/3611643.3616262}  have been proposed, they do not naturally support test reuse. Realizing this direction also requires addressing the substantial noise introduced by test harnesses in candidate clone pairs (Finding 4), which complicates identifying reusable tests. Future work can enhance these techniques to facilitate test reuse while mitigating such noise.

\noindent\textbf{Lightweight semantic filtering for traditional detectors.} Our evaluation shows that LLM detectors are more robust across code types, whereas traditional detectors substantially degrade on test code (Finding 6). However, LLM-based detection requires class-level pre-filtering for repository-scale inputs and remains costly even after filtering (Section~\ref{subsec:prefilt}). SAS-enhanced traditional detectors can approach or surpass LLM detectors (Findings 7 and 9). Thus, lightweight semantic filtering offers a practical way to enhance traditional detectors for code mapping detection when LLM-based inference is too costly.

\noindent\textbf{Importance of naming information.} Our evaluation shows that SAS, which captures semantic information embedded within names and documentation, helps improve clone detection for redesigned projects.  Among the four rules, identifier-name matching (R2) has the broadest impact on candidate rankings (Finding 8), highlighting the importance of naming information in code mapping detection. This provides empirical evidence for future research on identifier renaming in redesigned projects.

% Among all information, Finding 8 shows that identifier names is the effective in finding code mappings. This highlights the importance of naming information, and provides empirical evidence for future research on identifier renaming in redesigned projects. 
\section{Conclusion}
This paper investigates software reuse in cross-repository redesigns through an action research–driven, bidirectional study of the ongoing \soot/\sootup redesign. Our study shows that redesign is non-linear and often requires bidirectional reuse, while tests and residual bugs are frequently overlooked during migration. We identify tracking corresponding code and tests as a key challenge and address it by retrofitting existing clone detectors with a post-filtering method based on Semantic Alignment Scores (SAS). Evaluation across three redesign cases (\soot/\sootup, \fbugs/\sbugs, and ANTLR3/ANTLR4) shows that SAS filtering substantially improves traditional detectors, with gains of up to 0.62 in precision and 0.34 in average F1, enabling strong traditional detectors to approach or surpass the CM-detection effectiveness of LLM detectors without LLM inference costs. We further contribute a benchmark of \numdataset labeled method pairs for code mapping research, while our open-source contributions demonstrate the practical utility of the approach. These contributions provide actionable support for software reuse in ongoing redesign projects.

\noindent \textbf{Data Availability}
Our code, annotation codebook, and experimental results are available online in our replication package~\citep{replication}.
% Our implementation, data, and scripts are available at 
% \url{https://doi.org/10.5281/zenodo.19324267},
% with the analyzed project versions being \soot~\sversion, \sootup~\stversion, \sbugs~\sbversion, and \fbugs~\fbversion.

\bibliographystyle{spbasic}      % basic style, author-year citations
\bibliography{refs}   % name your BibTeX data base

\newpage
\appendix

\renewcommand{\thetable}{A.\arabic{table}}
\setcounter{table}{0}

\section{Effectiveness Results for the GC Objective}

% \begin{adjustbox}{angle=-90,center}%
\rotatebox{-90}{%
\begin{talltblr}[
    theme = mockcaption,
    caption = {Effectiveness of clone detectors before and after SAS filtering for the GC objective.},
    label = {tab:baselines-gc},
    remark{Format} = {``X/Y/Z'' denotes the original result, filtered result, and change ($Y-X$), respectively.},
]{
    width=0.9\textheight,
    colspec = {l l *{8}{Q[r]}},
    cells = {font=\scriptsize, m},
    colsep = 3pt, % 调整列间距，列宽=文字内容的宽度+2*colsep
    rowsep = 1pt, 
    % % --- merge rows by pair ---
    cell{3}{1} = {r=6}{m}, % soot/sootup
    cell{9}{1} = {r=6}{m},
    cell{15}{1} = {r=6}{m},
    cell{21}{1} = {r=6}{m, font=\bfseries\scriptsize}, % all pairs
    % % % --- color ---
    row{even} = {bg=gray!10}, 
    column{1} = {bg=white},
    % % --- 线条控制 ---
    hline{1, 3, 21, 27} = {1.2pt, solid}, % 顶线
    hline{2, 9, 15} = {solid},
    vline{7} = {solid},
    % % --- header row ---
    cell{1}{3,7} = {c=4}{m},
    cell{1}{1,2} = {r=2}{m},
    row{1,2} = {halign=c, font=\bfseries\scriptsize, bg=gray!30},
}

Pair & Tool & Production Code & &     &    & Test Code &      &      &          \\
     &      &  FPR &  Prec. &  Rec. &  Avg. F1 &  FPR &  Prec. &  Rec. &  Avg. F1  \\

\soot/\sootup & NiCad & 0.27/\textbf{0.05}/-0.22 & 0.58/0.88/0.30 & 0.58/0.57/\textbf{-0.01} & 0.656/0.774/0.117 & 0.01/0.00/-0.01 & 0.00/0.00/0.00 & 0.00/0.00/0.00 & 0.497/\textbf{0.500}/0.003 \\
\soot/\sootup & DeepSim & \textbf{0.39}/0.01/-0.39 & 0.38/0.96/\textbf{0.58} & 0.38/0.33/-0.05 & 0.492/0.653/\textbf{0.161} & \textbf{0.56}/\textbf{0.01}/-0.55 & 0.00/0.00/0.00 & 0.00/0.00/0.00 & 0.306/0.499/\textbf{0.193} \\
\soot/\sootup & CCStokener & 0.35/\textbf{0.05}/-0.30 & 0.47/0.85/0.38 & 0.49/0.49/\textbf{-0.01} & 0.570/0.726/0.157 & 0.04/0.00/-0.04 & 0.00/0.00/0.00 & 0.00/0.00/0.00 & 0.491/\textbf{0.500}/0.009 \\
\soot/\sootup & SSCD & 0.06/0.03/-0.03 & 0.91/0.95/0.04 & \textbf{0.88}/\textbf{0.83}/-0.05 & \textbf{0.913}/\textbf{0.908}/-0.004 & 0.48/0.00/-0.48 & 0.00/0.00/0.00 & 0.00/0.00/0.00 & 0.344/\textbf{0.500}/0.156 \\
\soot/\sootup & GPT-OSS & 0.07/0.01/-0.07 & 0.86/0.98/0.12 & 0.71/0.62/-0.10 & 0.828/0.821/-0.007 & 0.00/0.00/0.00 & 0.00/0.00/0.00 & 0.00/0.00/0.00 & \textbf{0.500}/\textbf{0.500}/0.000 \\
\soot/\sootup & Qwen-Coder & 0.00/0.00/0.00 & \textbf{1.00}/\textbf{1.00}/0.00 & 0.38/0.37/\textbf{-0.01} & 0.692/0.687/-0.004 & 0.00/0.00/0.00 & 0.00/0.00/0.00 & 0.00/0.00/0.00 & \textbf{0.500}/\textbf{0.500}/0.000 \\
\fbugs/\sbugs & NiCad & 0.02/0.02/0.00 & 0.99/0.99/0.00 & 0.78/0.77/0.00 & 0.831/0.829/-0.002 & 0.00/0.00/0.00 & \textbf{1.00}/\textbf{1.00}/0.00 & 0.32/0.30/-0.02 & 0.529/0.515/-0.014 \\
\fbugs/\sbugs & DeepSim & \textbf{0.76}/0.06/-0.70 & 0.58/0.95/\textbf{0.37} & 0.47/0.47/0.00 & 0.362/0.617/\textbf{0.255} & \textbf{0.82}/0.05/-0.77 & 0.27/0.86/\textbf{0.59} & 0.14/0.14/0.00 & 0.153/0.367/\textbf{0.214} \\
\fbugs/\sbugs & CCStokener & 0.30/\textbf{0.10}/-0.19 & 0.86/0.95/0.08 & 0.86/0.84/-0.01 & 0.776/0.846/0.069 & 0.21/\textbf{0.16}/-0.06 & 0.88/0.91/0.02 & 0.74/0.69/-0.05 & 0.736/0.730/-0.007 \\
\fbugs/\sbugs & SSCD & 0.07/0.05/-0.02 & 0.97/0.98/0.01 & \textbf{0.99}/\textbf{0.98}/-0.01 & \textbf{0.962}/\textbf{0.965}/0.003 & 0.05/0.05/0.00 & 0.98/0.98/0.00 & \textbf{0.95}/\textbf{0.91}/-0.05 & \textbf{0.947}/\textbf{0.914}/-0.033 \\
\fbugs/\sbugs & GPT-OSS & 0.02/0.00/-0.02 & 0.99/\textbf{1.00}/0.01 & 0.84/0.84/0.00 & 0.873/0.879/0.006 & 0.01/0.00/-0.01 & 0.99/\textbf{1.00}/0.01 & 0.44/0.44/0.00 & 0.617/0.620/0.003 \\
\fbugs/\sbugs & Qwen-Coder & 0.00/0.00/0.00 & \textbf{1.00}/\textbf{1.00}/0.00 & 0.79/0.79/0.00 & 0.847/0.847/0.000 & 0.00/0.00/0.00 & \textbf{1.00}/\textbf{1.00}/0.00 & 0.25/0.25/0.00 & 0.480/0.476/-0.004 \\
\alrt/\alrf & NiCad & 0.00/0.00/0.00 & \textbf{1.00}/\textbf{1.00}/0.00 & 0.26/0.21/-0.04 & 0.663/0.633/-0.030 & 0.10/0.01/-0.10 & 0.62/\textbf{0.97}/0.35 & 0.45/0.45/0.00 & 0.689/0.761/0.072 \\
\alrt/\alrf & DeepSim & \textbf{0.62}/0.00/-0.62 & 0.12/\textbf{1.00}/\textbf{0.88} & 0.36/0.30/-0.06 & 0.337/0.691/\textbf{0.354} & \textbf{0.79}/0.02/-0.77 & 0.29/0.94/\textbf{0.65} & 0.88/0.86/-0.01 & 0.386/0.932/\textbf{0.547} \\
\alrt/\alrf & CCStokener & 0.42/\textbf{0.05}/-0.38 & 0.31/0.74/0.43 & 0.81/\textbf{0.55}/-0.26 & 0.582/0.780/0.198 & 0.54/\textbf{0.03}/-0.51 & 0.39/0.93/0.54 & \textbf{0.92}/\textbf{0.89}/-0.03 & 0.584/\textbf{0.938}/0.354 \\
\alrt/\alrf & SSCD & 0.23/0.02/-0.21 & 0.48/0.90/0.42 & \textbf{0.87}/\textbf{0.55}/-0.32 & \textbf{0.736}/\textbf{0.813}/0.077 & 0.09/0.01/-0.08 & 0.79/\textbf{0.97}/0.18 & 0.86/0.85/-0.01 & \textbf{0.877}/0.936/0.060 \\
\alrt/\alrf & GPT-OSS & 0.02/0.00/-0.02 & 0.82/\textbf{1.00}/0.18 & 0.30/0.19/-0.11 & 0.676/0.617/-0.060 & 0.02/0.02/0.00 & 0.94/0.93/-0.01 & 0.62/0.55/-0.07 & 0.835/0.802/-0.033 \\
\alrt/\alrf & Qwen-Coder & 0.00/0.00/0.00 & \textbf{1.00}/\textbf{1.00}/0.00 & 0.17/0.17/0.00 & 0.600/0.600/0.000 & 0.01/0.01/0.00 & \textbf{0.97}/\textbf{0.97}/0.00 & 0.48/0.48/0.00 & 0.776/0.776/0.000 \\
All Pairs & NiCad & 0.12/0.02/-0.10 & 0.82/0.96/0.13 & 0.66/0.65/-0.01 & 0.773/0.817/0.044 & 0.05/0.00/-0.04 & 0.82/\textbf{0.99}/0.17 & 0.35/0.34/-0.01 & 0.649/0.664/0.015 \\
All Pairs & DeepSim & \textbf{0.55}/0.02/-0.54 & 0.40/0.96/\textbf{0.56} & 0.43/0.41/-0.03 & 0.439/0.681/\textbf{0.243} & \textbf{0.71}/0.02/-0.69 & 0.22/0.90/\textbf{0.68} & 0.32/0.32/-0.01 & 0.299/0.640/\textbf{0.341} \\
All Pairs & CCStokener & 0.36/\textbf{0.06}/-0.30 & 0.63/0.91/0.28 & 0.73/0.70/-0.03 & 0.680/0.822/0.142 & 0.28/\textbf{0.04}/-0.24 & 0.63/0.91/0.28 & 0.78/0.74/-0.04 & 0.736/0.860/0.124 \\
All Pairs & SSCD & 0.12/0.03/-0.09 & 0.87/0.96/0.09 & \textbf{0.94}/\textbf{0.89}/-0.05 & \textbf{0.907}/\textbf{0.932}/0.025 & 0.22/0.01/-0.20 & 0.73/0.97/0.25 & \textbf{0.93}/\textbf{0.89}/-0.04 & \textbf{0.837}/\textbf{0.946}/0.109 \\
All Pairs & GPT-OSS & 0.04/0.00/-0.04 & 0.94/0.99/0.05 & 0.74/0.70/-0.04 & 0.856/0.854/-0.001 & 0.01/0.01/0.00 & 0.97/0.98/0.01 & 0.49/0.47/-0.02 & 0.753/0.744/-0.008 \\
All Pairs & Qwen-Coder & 0.00/0.00/0.00 & \textbf{1.00}/\textbf{1.00}/0.00 & 0.59/0.59/0.00 & 0.800/0.799/-0.001 & 0.00/0.00/0.00 & \textbf{0.99}/\textbf{0.99}/0.00 & 0.31/0.31/0.00 & 0.646/0.644/-0.002 \\
\end{talltblr}
}
% \end{adjustbox}

\end{document}